\documentclass[12pt]{article}
\usepackage{times,rjscmplx,apalike}
\usepackage{graphicx,pifont,wrapfig,amsmath,amssymb,amsfonts}
\title{The case of the trapped singularities}

\author{R. Ball}

\address{Department of Theoretical Physics, 
Research School of Physical Sciences \& Engineering,
The Australian National University,
Canberra ACT 0200 AUSTRALIA
}
\email{Rowena.Ball@anu.edu.au}

\begin{document}
\maketitle

\begin{abstract}

A case study in bifurcation and stability analysis is presented, in which reduced dynamical system modelling yields substantial new global and predictive information about the behaviour of a complex system. The first smooth pathway, free of pathological and persistent degenerate singularities, is surveyed through the parameter space of a nonlinear dynamical model for a gradient-driven, turbulence--shear flow energetics in magnetized fusion plasmas. Along the route various obstacles and features are identified and treated appropriately. An organizing centre of low codimension is shown to be robust, several trapped singularities are found and released, and domains of hysteresis, threefold stable equilibria, and limit cycles are mapped. Characterization of this rich dynamical landscape achieves unification of previous disparate models for plasma confinement transitions, supplies valuable intelligence on the big issue of shear flow suppression of turbulence, and suggests targeted experimental design, control and optimization strategies.

\end{abstract}

\section{Introduction}

The collective dynamics of complex distributed systems often can be usefully described in terms of a superposition
of rate processes or frequencies which determine the changes in  
macroscopically
measurable variables as energy flows through the system; that is, a dynamical model expressed as a system of coupled ordinary differential equations in a few averaged  state variables or mode coefficients
and several, independently tunable, parameters
that represent physical properties or external controls. This type of reduced (or low-order or low-dimensional)  modelling averages over 
space, mode spectrum structure, single-particle
dynamics and other details, 
but the payoff lies in its amenity to sophisticated 
analytic theory and methods that enable us to track important 
qualitative features 
in the collective dynamics, such as singularities, bifurcations, and
stability changes, broadly over the parameter space.

Motivated by the need for improved guidance and control of the (mostly bad) behaviour of fusion plasmas in magnetic containers, I elaborate in this work a case study in bifurcation and stability analysis in which reduced dynamical system modelling yields new global and predictive information about gradient driven turbulence--flow energetics that is complementary to direct numerical simulation and can guide experimental design. 

Reduced dynamical models 
are powerful tools for describing and analysing complex systems such as turbulent plasmas and fluids, primarily because they are supported 
by well-developed mathematics that gives qualitative and global insight, such as singularity, bifurcation, stability, and symmetry theory.
In principle one can map analytically the bifurcation structure of the entire phase and parameter space of a reduced dynamical system, but this feat is not possible for an infinite-dimensional system, or partial differential equations,
and not practicable for systems of high order\footnote{By ``reduced'' I mean having a 
phase space dimension of less than four or five, the most influential
 realm of qualitative theories.}.
 The usefulness of such models seems to be no coincidence, too: in turbulent systems generally, which in detail are both complex and complicated, the dynamics seems to take place in a low-dimensional subspace \cite{Holmes:1996}.

 It seems paradoxical that enthusiasm for low-dimensional 
modelling and qualitative analysis of fluid and plasma systems has paced 
the ever larger direct numerical simulations of their flow fields. This is an exemplar  of how the simplexity and complicity \cite{Cohen:1994} juxtaposition can work well: these methods affirm each other, for both have common ground in 
the universal conservation equations for fluid flow (as well as separate 
bases in mathematics and computational science).
Developments in one feed developments in~the~other.  Reduced dynamical models can give 
insights into the physics and dynamics of a system in a way that 
is complementary to brute-force numerical simulations of the detailed, 
spatially distributed models from which they are derived. 
In practice this complementarity means that
low-order models (which capture few or no spatial modes) can be used 
to channel information gleaned from the generic, qualitative
structure of the parameter space --- attractors, critical points of onset, 
stability properties, and so on --- to numerical simulations 
(which contain all spatial modes but, on their own, bring little 
physical understanding), giving them purpose and meaning. 
In turn the fluid simulations serve as virtual experiments to validate
the low-order approach.

It is reasonable, therefore, to assert that improved 
low-dimensional dynamical models for plasmas and fluids could provide 
numerical experimenters with new and interesting challenges that will
continue to push the limits of computational science and technology.

\subsection{Two-dimensional fluid motion in plasmas}

Fusion plasmas in magnetic containers, such as those in tokamak or stellarator experiments, are strongly driven nonequilibrium systems in which the kinetic energy of small-scale turbulent fluctuations can drive the formation of large-scale coherent structures such as shear and zonal flows. This inherent tendency to self-organize is a striking characteristic of flows where Lagrangian fluid elements see a  predominantly two-dimensional  velocity field, and is a consequence of the inverse energy cascade \cite{Kraichnan:1980}. The distinctive properties of quasi two-dimensional fluid motion are the basis of natural phenomena such as zonal and coherent structuring of planetary flows, but are generally under-exploited in technology. 

In plasmas the most potentially useful effect of two-dimensional fluid motion is  suppression of high wavenumber turbulence that generates cross-field transport fluxes and degrades confinement \cite{Terry:2000}. Suppression of turbulent transport can manifest temporally as a spontaneous and more-or-less abrupt enhancement of sheared poloidal or zonal flows and concomitant damping of density fluctuations, and spatially as the rapid development of a localized transport barrier or steep density gradient. The phenomenon is often called low- to high-confinement (L--H) transitions and has been the subject of intensive experimental, \textit{in numero,} and theoretical and modelling investigations since the 1980s.

\subsection{Confinement transitions and reduced dynamical systems}

The large and lively primary literature on reduced dynamical models for  confinement transitions and associated oscillations in plasmas represents a sort of consensus on the philosophy behind qualitative analysis, if not on the details of the models themselves.
What motivates this approach is the predictive 
power that a unified, low-order description of the macroscopic dynamics would have in the management of confinement states. Since it is widely acknowledged that control of turbulent transport is
crucial to the success of the world-wide fusion energy program \cite{doe1:2000,Fujisawa:2000}
it is important to develop predictive models for efficient management of
access to, and sustainment of, high confinement r\'egimes. For example, 
if one plans to maintain a high confinement state 
 at a relatively low power input by exploiting the hysteresis in the transition
 it would be useful, not to mention cheaper, to know in advance which 
parameters control
the shape and extent of hysteresis, or whether it can exist at all in the 
operating space of a particular system, or
whether a transition  will be oscillatory. 

However, it has been shown \cite{Ball:2000,Ball:2001,Ball:2002} that many of the models in the literature are structurally flawed.  They often
contain pathological or persistent degenerate (higher order) singularities. An associated issue is that of overdetermination, where  near  a persistent degenerate singularity there may be more defining equations than variables. Consequently much of the discussion in the literature concerning confinement transitions is qualitatively wrong. 
\textbf{Such models cannot possibly have predictive power}. 

The heart of the matter
lies in the mapping between the bifurcation structure and stability properties of a 
dynamical model and the physics of the process it is supposed
to represent: if we probe this relationship we
find that degenerate singularities ought to correspond to some essential physics (such as fulfilling a symmetry-breaking imperative, or the onset of hysteresis), or they are pathological. 
In the first case we can usually unfold the singularity 
in a physically meaningful way; 
in the other case we know that something 
is amiss and we should revise our assumptions. Degenerate
singularities are good because they provide opportunities to 
 improve a model and its predictive capabilities, but bad when they are not recognized as such. 
 
\subsection*{}

The literature on confinement transitions has two basic strands: (1) Transitions are an internal, quasi two-dimensional flow, phenomenon and occur spontaneously when the rate of upscale transfer of kinetic energy from turbulence to shear and zonal flows exceeds the nonlinear dissipation rate \cite{Diamond:1992}; (2) Transitions are due to nonambipolar ion orbit losses near the plasma edge, the resulting electric field providing a torque which drives the poloidal shear flow nonlinearly  \cite{Itoh:1988,Shaing:1989,Stringer:1993}.  
These two  different views of the physics behind confinement transitions are smoothly
reconciled for the first time in this work. 

A systematic methodology for characterizing the equilibria of dynamical systems involves finding and classifying high-order singularities then perturbing around them to explore and map the bifurcation landscape \cite{Golubitsky:1985}. Broadly, this paper is about applying singularity theory as a diagnostic tool while an impasto picture of confinement transition dynamics is compounded. The bare-bones model is presented in section \ref{two} In section \ref{three} the global consequences of local symmetry-breaking are explored, leading to the discovery of an organizing centre and trapped degenerate singularities. This leads in to section \ref{four} where I unfold a trapped singularity smoothly by introducing another layer that models the neglected physics of downscale energy transfer. Section \ref{five} follows the qualitative changes to the bifurcation and stability structure that are due to potential energy dissipative losses. In section \ref{six} the unified model is presented, in which is included a direct channel between gradient potential energy and shear flow kinetic energy. 
The results and conclusions are summarized in section \ref{seven}

\section{\label{two} The basic model comprises three energy subsystems}

\begin{figure}[hc]
\begin{center}
\hbox{
\includegraphics[scale=0.5]{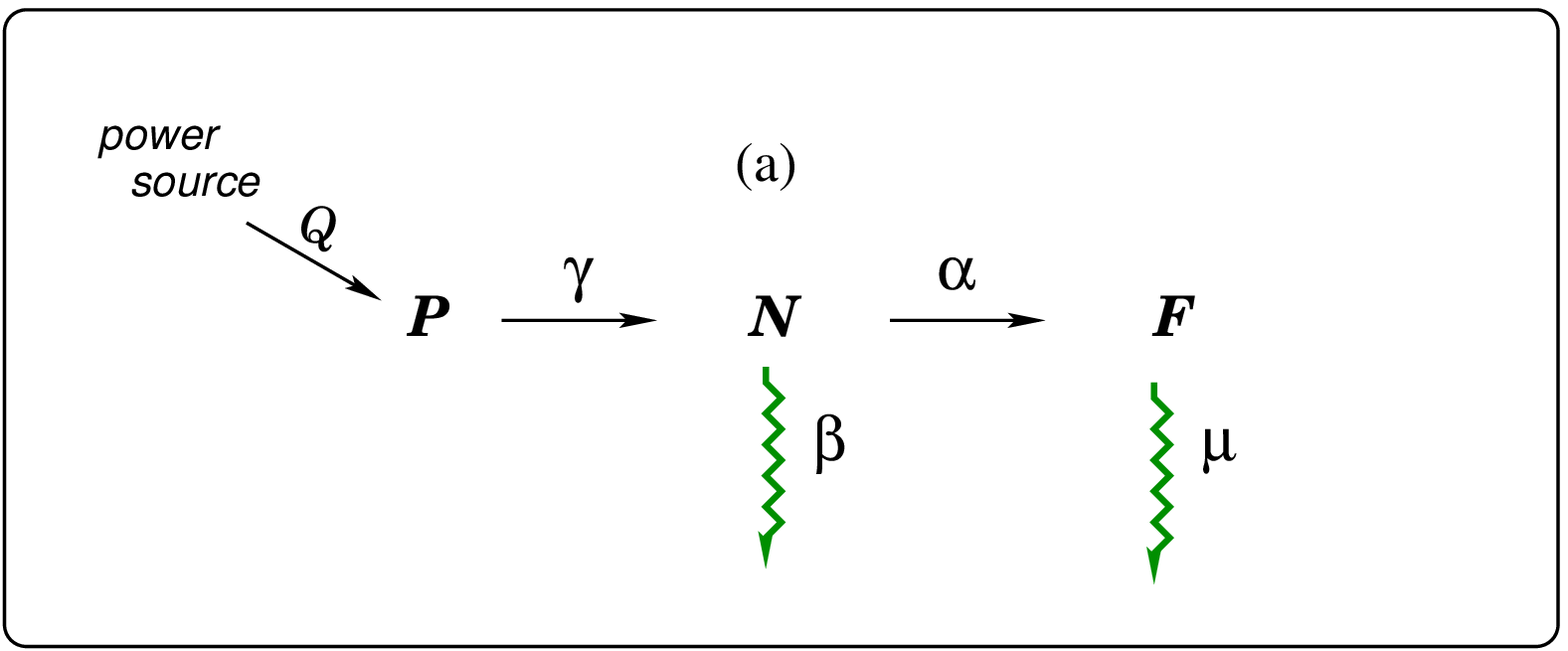}
\includegraphics[scale=0.5]{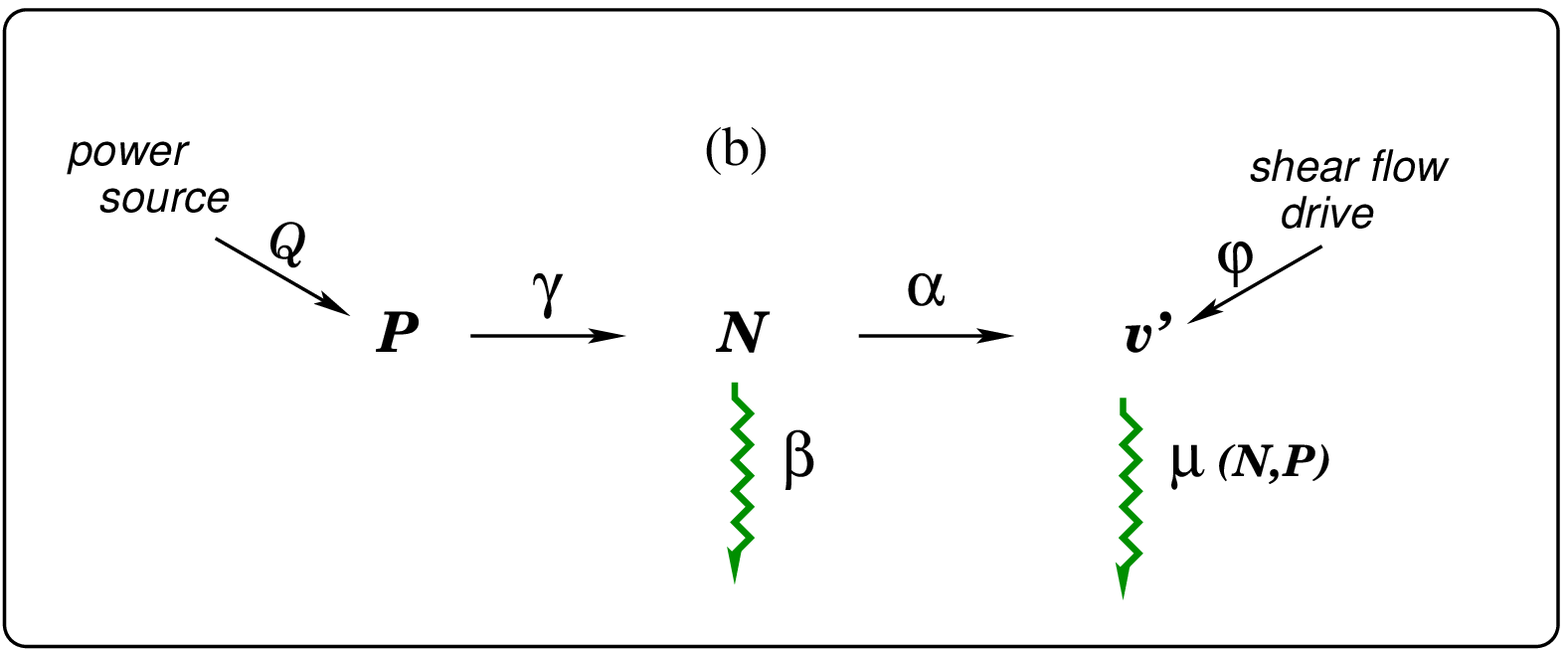}
}
\hbox{
\includegraphics[scale=0.5]{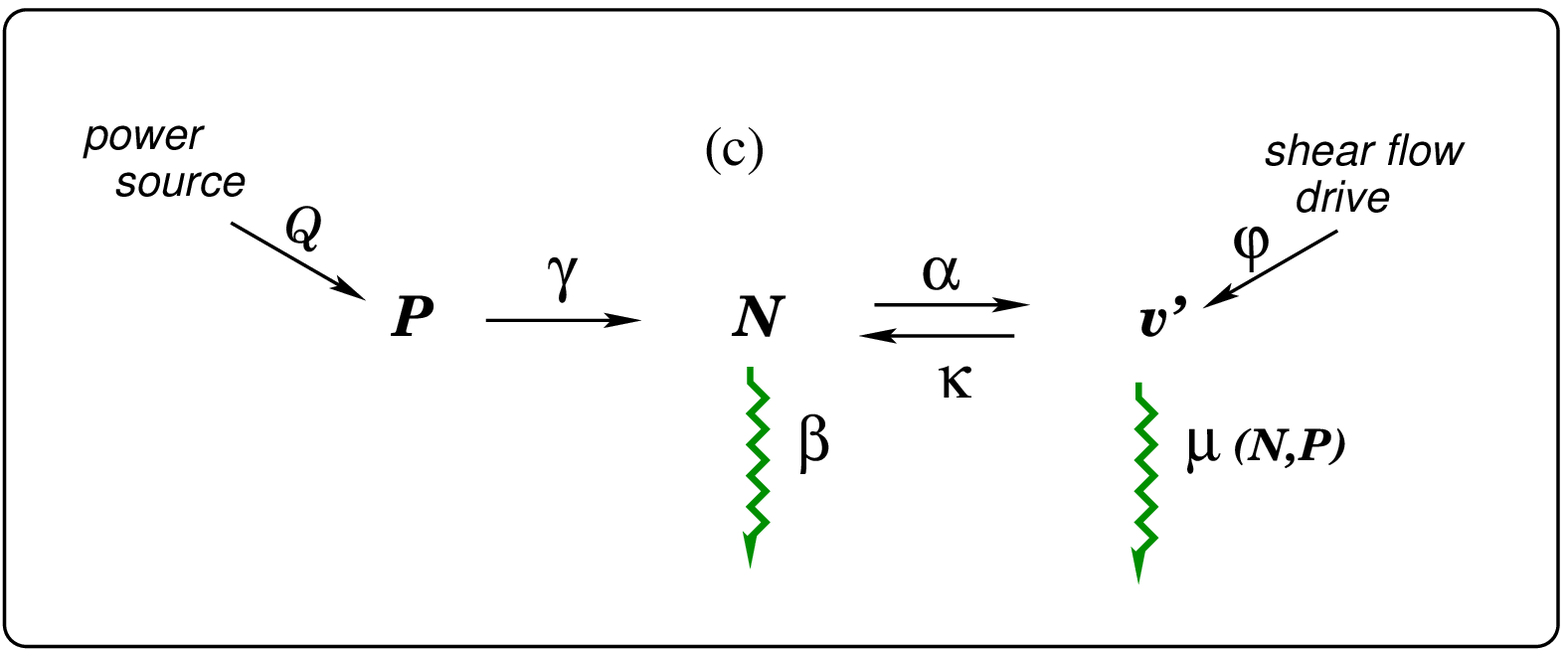}
\includegraphics[scale=0.5]{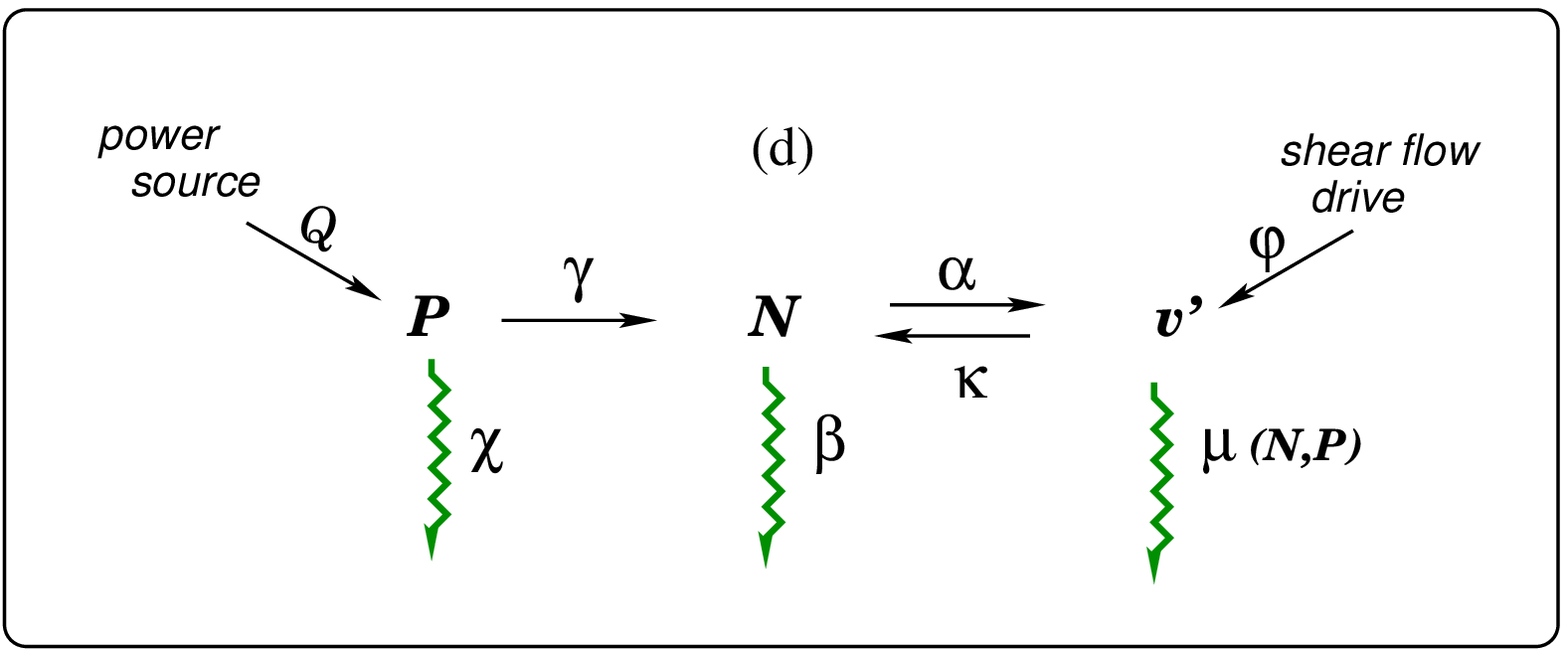}
}
\includegraphics[scale=0.5]{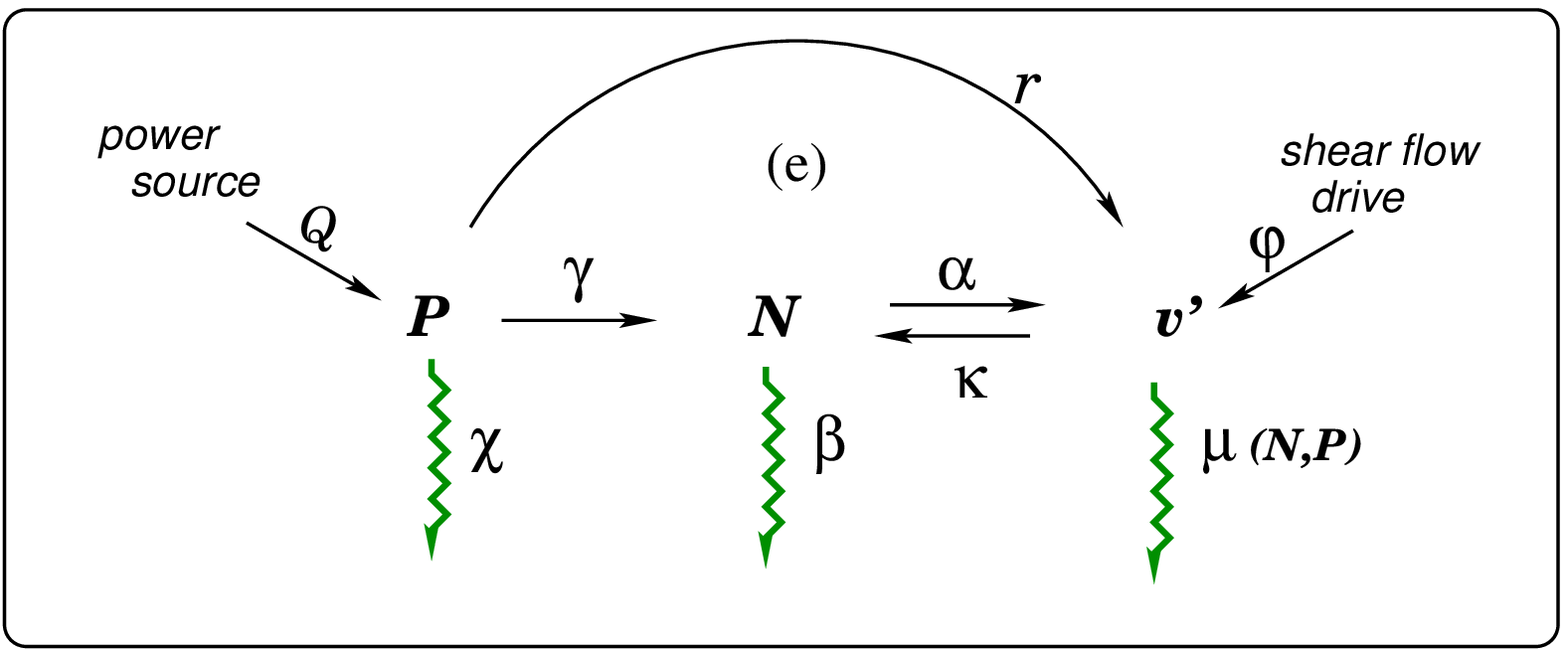}
\caption{\label{fig1} Energy transfer diagrams for the gradient-driven plasma turbulence--shear flow system. Annotated arrows denote rate processes; curly arrows indicate dissipative channels, straight arrows indicate inputs and transfer channels between the energy-containing subsystems. See text for explanations of each subfigure. }
\end{center}
\end{figure}

In the edge region of a plasma confinement experiment such as a tokamak or stellarator potential energy is stored in a steep pressure gradient which is fed by a power source near the centre. Gradient potential energy $P$ is converted to turbulent kinetic energy $N$, which is drawn off into stable shear flows, with kinetic energy $F$, and dissipation channels.  
The energetics of this simplest picture of confinement transition dynamics  are schematized in Fig.~\ref{fig1}(a). (Nomenclature for the quantities here and in the rest of the paper is defined in  Table \ref{tab1}.)

A skeleton dynamical system for this overall process can be written down directly from Fig.~\ref{fig1}(a) by inspection:
\begin{equation}\label{e1}
\begin{aligned}
\varepsilon\frac{dP}{dt} & = Q - \gamma\left(P,N,F\right)\\
\frac{dN}{dt} & = \gamma\left(P,N,F\right) - \alpha\left(P,N,F\right) - \beta\left(P,N,F\right)\\
\frac{dF}{dt} & =  \alpha\left(P,N,F\right) - \mu\left(P,N,F\right). 
\end{aligned}
\end{equation}
The power input $Q$ is assumed constant and the energy transfer and dissipation rates generally may be functions of the energy variables. A more physics-based derivation of this system was outlined in Ball \citeyear(2002){Ball:2002}, in which averaged energy integrals were taken of momentum and pressure convection equations in slab geometry, using an electrostatic approximation to eliminate the dynamics of the magnetic field energy. Equations \ref{e1} are fleshed out by substituting specific rate-laws for the general rate expressions on the right hand sides:  
\begin{subequations}\label{e2}
\begin{align}
\varepsilon\frac{dP}{dt} & = Q - \gamma NP\label{e2a}\\ 
\frac{dN}{dt} & =  \gamma NP - \alpha v^{\prime 2}N - \beta N^2\label{e2b}\\
2\frac{dv^\prime}{dt} & =  \alpha v^\prime N - \mu(P,N) v^\prime \label{e2c}\\
\mu(P,N) = & bP^{-3/2} + a PN, \label{e2d}
\end{align}
\end{subequations}
where $F = \pm v^{\prime^2}$. 
The rate expressions in Eqs \ref{e2} were derived in Sugama and Horton \citeyear(1995){Sugama:1995} and Ball \citeyear(2002){Ball:2002} from semi-empirical arguments or given as ansatzes. (Rate-laws for bulk dynamical processes are not usually derivable purely from theory, and ultimately must be tested against experimental evidence.)

The rest of this paper is concerned with the  character of the equilibria of Eqs \ref{e2} and modifications and extensions to this system. We shall study the type, multiplicity, and stability of attractors, interrogate degenerate or pathological singularities where they appear, and classify and map the bifurcation structure of the system. In doing this we shall attempt to answer questions such as: Are Eqs 
\ref{e2} or modified versions a good --- that is, predictive --- model of the system? Does the model  adequately reflect the known phenomenology of confinement transitions in fusion plasmas?  What is the relationship between the bifurcation properties of the model and the physics of confinement transitions? 

\section{\label{three} Symmetry-breaking has local and global consequences} 

\subsection{How to dissolve a pitchfork}

\begin{figure}[hc]
\begin{center}
\hspace*{-4mm}
\hbox{
\includegraphics[scale=0.58]{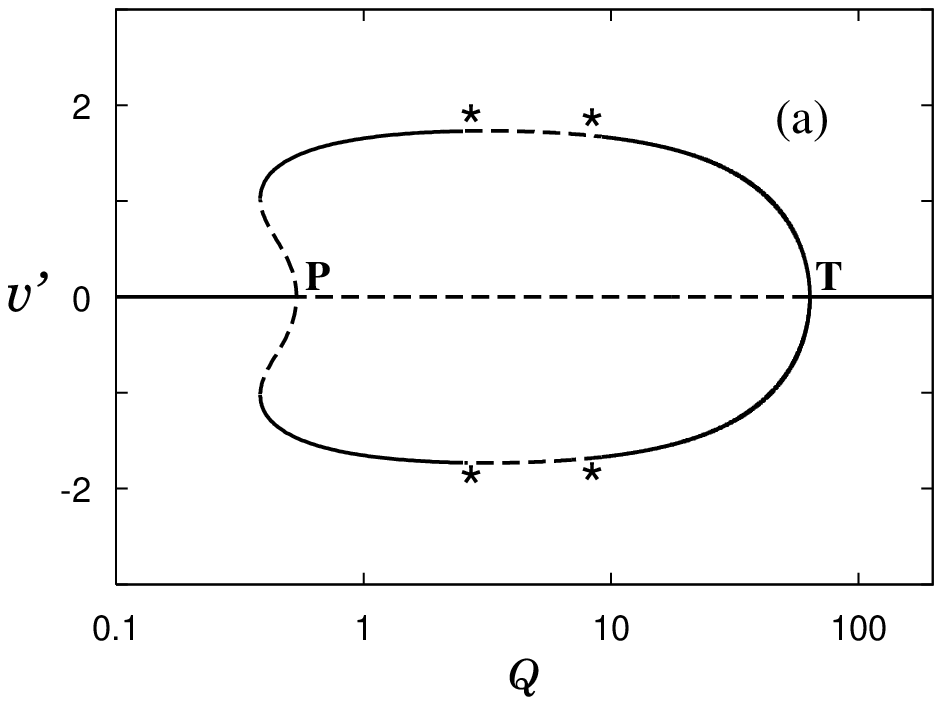}
\includegraphics[scale=0.58]{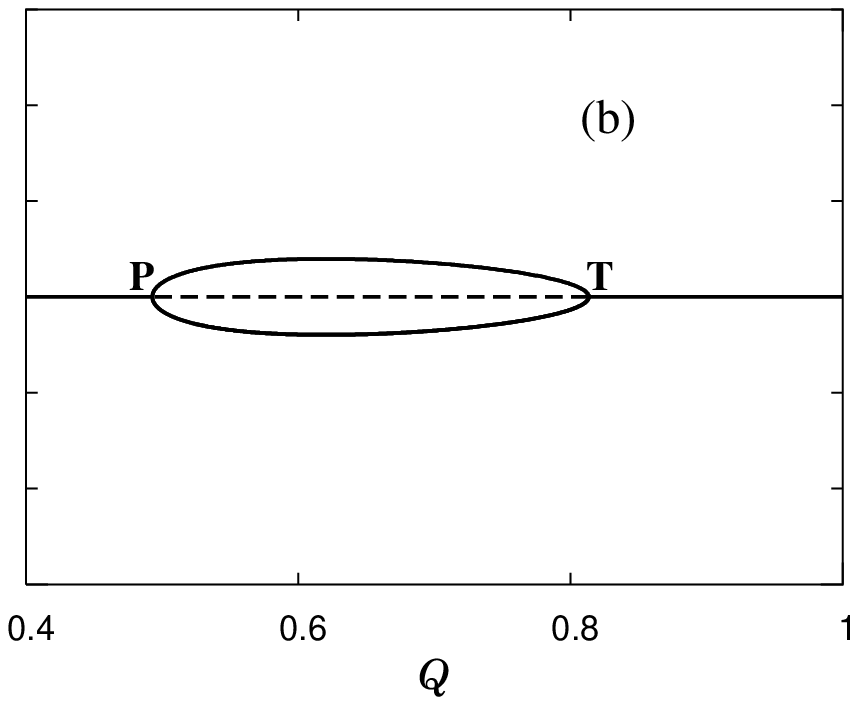}
\includegraphics[scale=0.58]{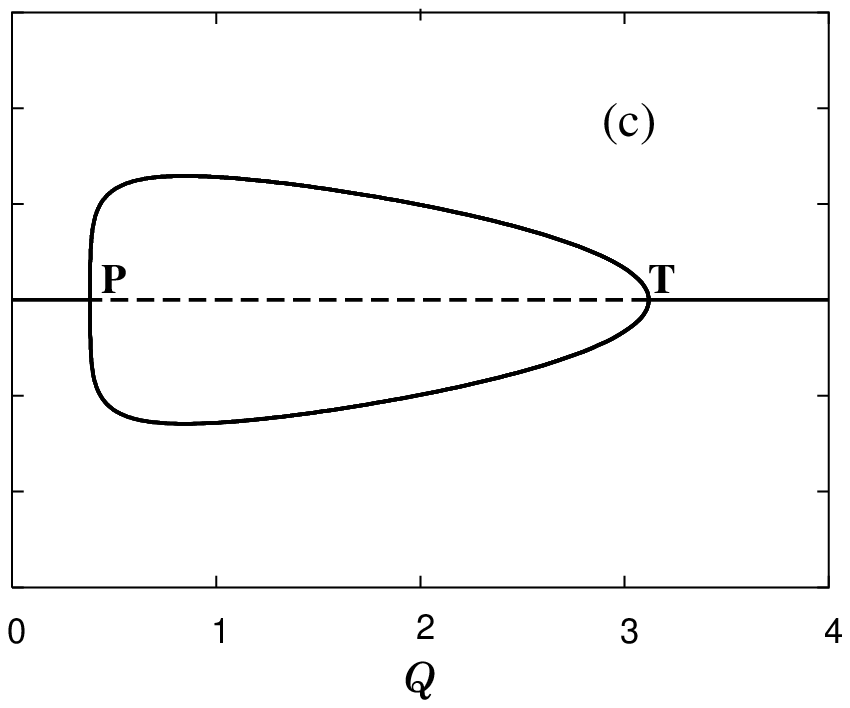}
}
\caption{\label{fig2} (a) $\beta = 1$,  (b) $\beta = 50$, (b) $\beta = \beta_{crit} \approx 18.58$. $\gamma =1$, $b=1$, $a=0.3$, $\alpha=2.4$, $\varepsilon=1.5$. }
\end{center}
\end{figure}
The equilibrium solutions of Eqs \ref{e2} are shown in the bifurcation diagrams of Fig. \ref{fig2}, where the shear flow $v^\prime$ is chosen as the state variable and the power input $Q$ is chosen as the principal bifurcation or control parameter. (In these and subsequent bifurcation diagrams stable equilibria are indicated by solid lines and unstable equilibria are indicated by dashed lines.)
Several bifurcation or singular points are evident. 

The four points in (a) annotated by asterisks, where the stability of solutions changes,  are Hopf bifurcations to limit cycles, which are discussed in section \ref{three-two}. 

On the line $v^\prime=0$ the singularity \textbf{P} is found to satisfy the defining and non-degeneracy conditions for a pitchfork,
\begin{equation}\label{e3}
G=G_\zeta =G_{\zeta\zeta}=G_\lambda = 0, \,\,
G_{\zeta\zeta\zeta}\neq 0, \,G_{\zeta\lambda}\neq 0, 
\end{equation}
where $G$ is the bifurcation equation derived from the zeros of Eqs \ref{e2}, $\zeta$ represents the chosen state variable, $\lambda$ represents the chosen control or principal bifurcation parameter, and the subscripts denote partial derivatives. In the qualitatively different bifurcation diagrams (a) and (b) the dissipative parameter $\beta$ is relaxed either side of the critical value given in (c), where the perfect, twice-degenerate pitchfork is represented. Thus for $\beta<\beta_\text{crit}$ (a poorly dissipative system) the turning points in (a) appear and the system may also show oscillatory behaviour. The dynamics are less interesting for $\beta>\beta_\text{crit}$ (a highly dissipative system) as in (b) because the turning points, and perhaps also the Hopf bifurcations, cannot occur.

However, \textbf{P} is persistent through variations in $\beta$ or any other parameter in Eqs \ref{e2}. (This fact was not recognized in some previous models for confinement transitions, where such points were wrongly claimed to represent second-order phase transitions.) 
Typically the pitchfork is associated with a fragile symmetry in the dynamics of the modelled physical system. The symmetry in this case is obvious from Fig. \ref{fig2}: in principle the shear flow can be in either direction equally. In real life (or \textit{in numero}), experiments are always subject to perturbations that determine a preferred direction for the shear flow, and the pitchfork is inevitably dissolved.   
In this case the perturbation is an effective force or torque from any asymmetric shear-inducing mechanism, such as friction with neutrals in the plasma or external sources, and acts as a shear flow driving rate. Assuming this rate to be small and independent of the variables over the characteristic timescales for the other rate processes in the system, we may revise the shear flow evolution Eq. \ref{e2c} as 
\begin{equation} \label{e4}
2\frac{dv^\prime}{dt}  =  \alpha v^\prime N - \mu(P,N) v^\prime + \varphi,
\end{equation}
where the symmetry-breaking term $\varphi$ models the shear flow drive. The corresponding energy transfer schematic is shown in Fig. \ref{fig1}(b).

The pitchfork \textbf{P} in Fig. \ref{fig2} (c) can now be obtained exactly by applying the conditions (\ref{e3}) to the zeros of Eqs \ref{e2a}, \ref{e2b}, and \ref{e4}, with \ref{e2d}, and
with $\zeta\equiv v^\prime$ and $\lambda\equiv Q$:
\begin{equation}\tag{\textbf{P}}
\left(v^\prime,Q,\beta,\varphi\right)=
\left(0,\frac{\alpha^2\gamma^2}{9a^2b},
\frac{2\alpha^3\gamma\sqrt{\alpha/a}}{27\sqrt{3}\:a^2b},0\right).
\end{equation}

The other singularity \textbf{T} on $v^\prime=0$ satisfies the defining and non-degeneracy conditions for a transcritical bifurcation, 
\begin{equation}\label{trans}
G=G_\zeta =G_\lambda = 0, \,\,
G_{\zeta\zeta}\neq 0, \, \,
\det\begin{pmatrix}G_{\zeta\zeta}& G_{\lambda\zeta}\\
                   G_{\lambda\zeta} & G_{\lambda\lambda}
\end{pmatrix}\equiv \det d^2 G < 0. 
\end{equation}
It is once-degenerate and also requires the symmetry-breaking parameter for exact definition.

\subsection{A walk along untrodden ways\label{three-two}}

\begin{figure}[ht]
\hspace*{-5mm}
\hbox{
\includegraphics[scale=0.83]{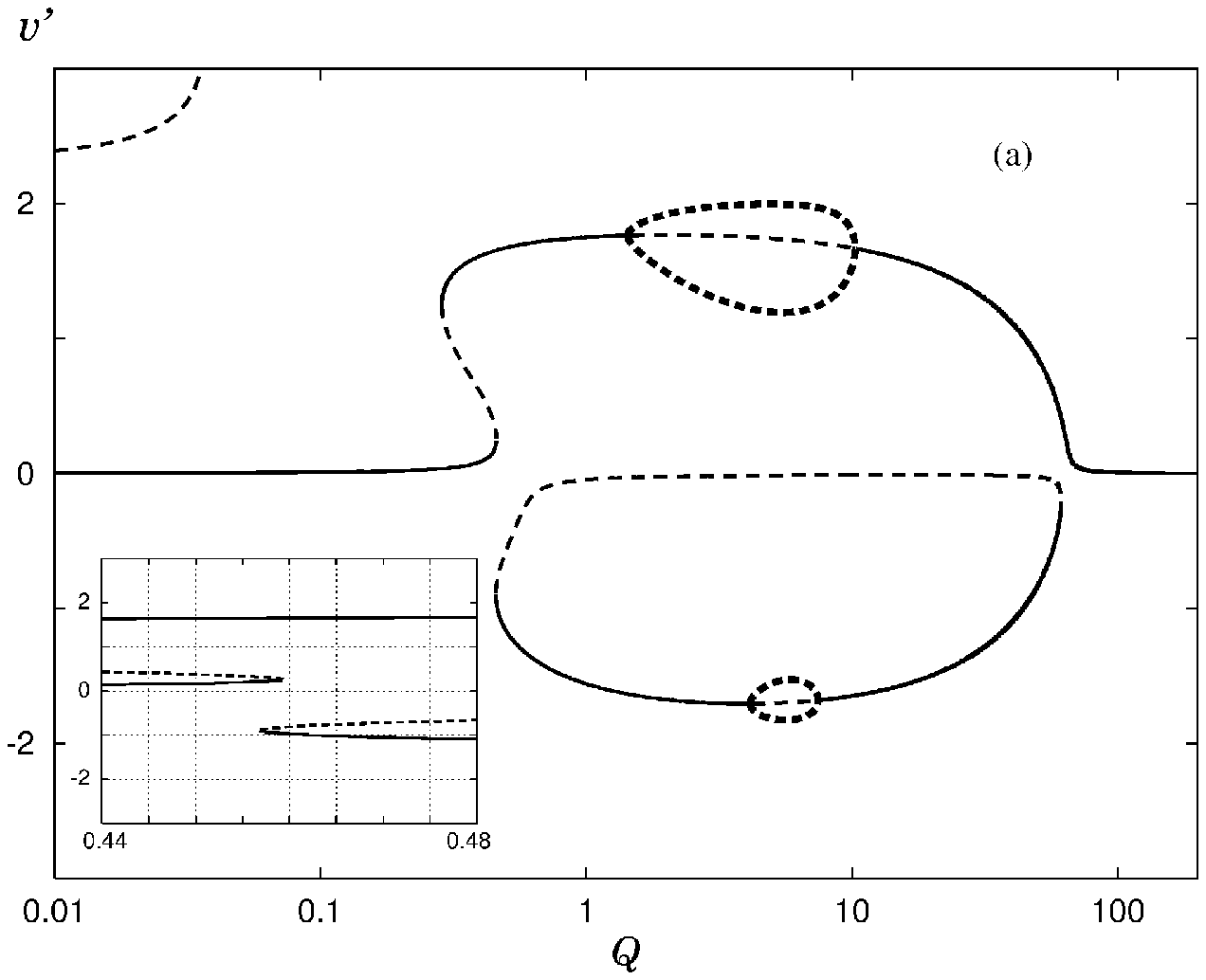}
\hspace*{-2mm}
\vbox{
\includegraphics[scale=0.78]{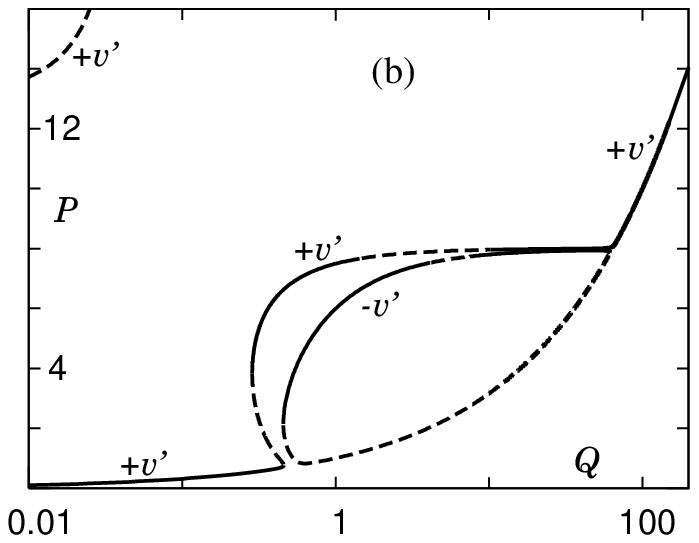}\\
\includegraphics[scale=0.78]{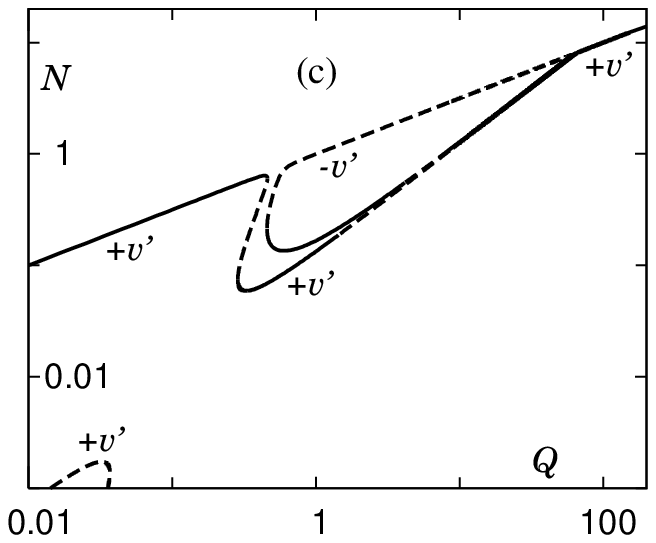}
}
}
\caption{\label{fig3} $\varphi=0.05$, other parameters as for Fig. \ref{fig2} (a). }
\end{figure}

A bifurcation diagram where \textbf{P} is fully unfolded, that is, for $\varphi\neq 0$ and $\beta \neq \beta_{crit}$, is shown in Fig. \ref{fig3}. This diagram is rich with information that speaks of the known and predicted dynamics of the system and of ways in which the model can be improved further, and which cannot be inferred or detected from the degenerate bifurcation diagrams of Fig. \ref{fig2}. It is worthwhile to step through Fig. \ref{fig3} in detail, with the energy schema Fig. \ref{fig1} (b) at hand.  Let us begin on the stable branch at $Q\sim 0.1$.

Here the pressure gradient is being charged up, the gradient potential energy is feeding the turbulence, and the shear flow is small but positive because the sign of the perturbation $\varphi$ is positive. As the power input $Q$ is increased quasistatically the shear flow begins to grow but at the turning point, where  solutions become unstable, there is a discontinuous transition to the upper stable branch in $v^\prime$ and $P$, (a) and (b), and to the lower stable branch in $N$, (c). At the given value of $\beta$ the hysteresis is evident: if we backtrack a little the back transition takes place at a lower value of $Q$. Continuing along the upper stable branch in $v^\prime$ we encounter another switch in stability; this time at a Hopf bifurcation to stable period one limit cycles. (In this and other diagrams the amplitude envelopes of limit cycle branches are marked by large solid dots.)  From the amplitude envelope we see that the oscillations grow as power is fed to the system then are extinguished rather abruptly, at a second Hopf bifurcation where the solutions regain stability. The shear flow decreases toward zero as the pressure-dependent anomalous viscosity, the second term in Eq. \ref{e2d}, takes over to dissipate the energy at high power input.  

The system may also be evolved to an equilibrium on the antisymmetric, $-v^\prime$ branch in Fig.~\ref{fig3}~(a), by choosing initial conditions\footnote{A note on initial conditions: The stable attractor that is spontaneously sought by a dynamical system will be the one for which the initial conditions lie within its basin of attraction. The overall bifurcation \textit{structure} of the type of dynamical system considered here is independent of initial conditions.}   appropriately or a large enough kick. However, if the power input then falls below the  turning point at $(v^\prime,Q)\approx (-0.9,0.46)$ we see an interesting phenomenon: the shear flow spontaneously reverses direction. The transient would nominally take the system toward the nearest stable attractor, the lower $+v^\prime$ branch, but since it would then be sitting very close to the lower $+v^\prime$ turning point small stochastic fluctuations could easily induce the transition to the higher $+v^\prime$ branch. See the inset zoom-in over this region in (a). Here is an example of a feature that is unusual in bifurcation landscapes, a domain over which there is fivefold multiplicity comprising three stable and two unstable equilibria. Two more examples of threefold stable domains will be shown in section \ref{five} 

The same equilibria depicted using $P$ and $N$ as dynamical variables in (b) and (c) are annotated to indicate whether they correspond to the $+v^\prime$ or $-v^\prime$ domain. For clarity the amplitude envelopes of the limit cycle solutions are omitted from (b) and (c). 

In the remainder of this paper I concentrate on the $+v^\prime$ branches and ignore the $-v^\prime$ domain.

\subsection{A metamorphosis of the dynamics\label{3.3}}

\begin{figure}[hc]
\hspace*{5mm}\hbox{
\includegraphics[scale=0.9]{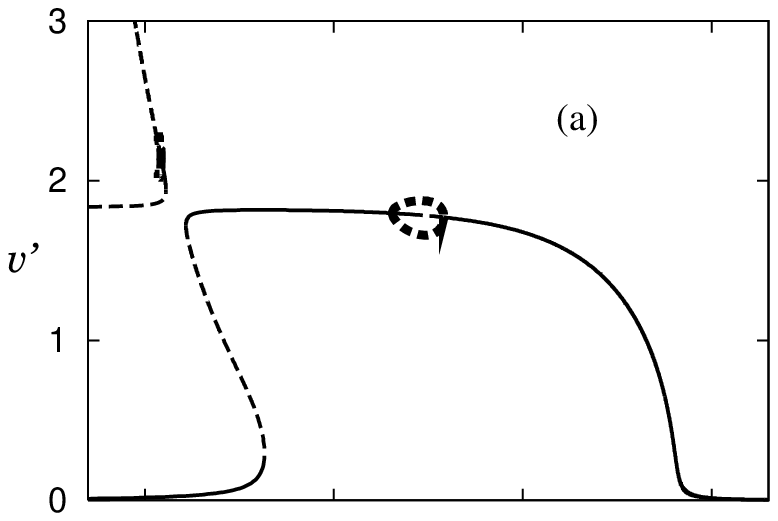}
\includegraphics[scale=0.9]{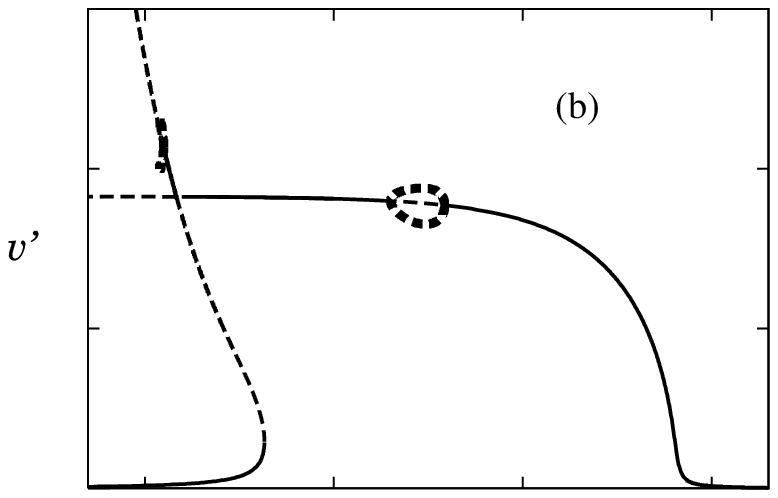}
}
\hspace*{5mm}\hbox{
\includegraphics[scale=0.9]{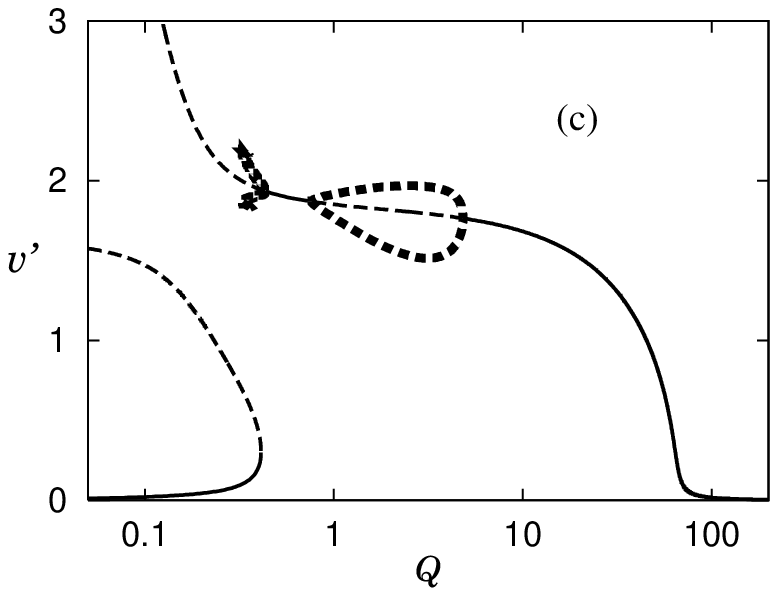}
\includegraphics[scale=0.9]{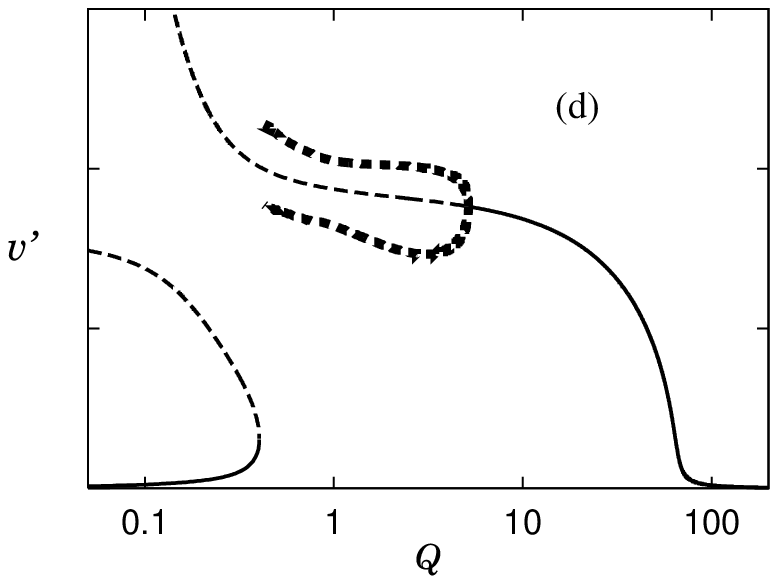}
}
\caption{\label{fig4} (a) $\varphi = 0.08$, (b) $\varphi \approx 0.08059 = \varphi_{Tm}$, (c) $\varphi = 0.1$, (d) $\varphi = 0.11$. $\varepsilon=1$, other parameters as for Fig. \ref{fig2} (a). }
\end{figure}

Now we approach the very heart of the model, the organizing centre; strangely enough via the branch of \emph{unstable} solutions that is just evident in Fig. \ref{fig3} (a) and (b) in the top left-hand corner and (c) in the lower left hand corner. The effects of symmetry-breaking are more far-reaching than merely providing a local universal unfolding of the pitchfork, for this
branch of equilibria was trapped as a singularity at $(v^\prime,Q)= (\infty,0)$ for $\varphi=0$. 
The organizing centre itself, described as a metamorphosis in Ball \citeyear(2002){Ball:2002}, can be encountered by varying $\varphi$. The sequence in Fig.~\ref{fig4} tells the story visually.

 The ``new'' unstable branch develops a Hopf bifurcation at the turning point. (Strictly speaking, this is a degenerate Hopf bifurcation, called DZE, where a pair of complex conjugate eigenvalues have zero real and imaginary components.) As $\varphi$ is tuned up to 0.08 (a) a segment of stable solutions becomes apparent as the Hopf bifurcation moves away from the turning point; 
the associated small branch of limit cycles can also just be seen. At $\varphi_{TM}$, the metamorphosis, the ``new'' and ``old'' branches exchange arms, (b). 

The metamorphosis satisfies the conditions (\ref{trans}) and is therefore an unusual, non-symmetric, transcritical bifurcation. 
It signals a profound change in the \textit{type} of dynamics that the system is capable of. For $\varphi>\varphi_{TM}$ a transition must still occur at the lower limit point, but there is no classical hysteresis, (c) and (d). In fact classical hysteresis is (locally) forbidden by the non-degeneracy condition $G_{\zeta\zeta}\neq 0$ in Eqs \ref{trans}. Various scenarios are possible in this r\'egime, including a completely non-hysteretic transition, a forward transition to a stable steady state and a back transition from a large period limit cycle, or forward and back transitions occurring to and from a limit cycle.

\subsection{The story so far}
The symmetry-broken model comprises Eqs \ref{e2a}, \ref{e2b}, and \ref{e4}, with \ref{e2d}. The bifurcation structure, some of which is depicted in Figs \ref{fig3} and \ref{fig4}, predicts various behaviours:
\begin{itemize}
\item Shear flow suppression of turbulence;
\item Smooth, hysteretic, non-hysteretic, and oscillatory transitions;
\item Spontaneous and kicked reversals in direction of shear flow; 
\item Saturation then decrease of the shear flow with power input due to pressure-dependent anomalous viscosity;  
\item A metamorphosis of the dynamics through a transcritical bifurcation.
 \end{itemize}
A critical appraisal of experimental evidence that supports the qualitative structure of this model is given in Ball \citeyear(2004){Ball:2004a}. With the exception of the last item all of the above dynamics have been observed in magnetically contained fusion plasma systems. The model would therefore seem to be a ``good'' and ``complete'' one, in the sense of being free of pathological or persistent degenerate singularities and reflecting observed behaviours. 

However, there are several outstanding issues that suggest the model is still incomplete. One issue arises as a gremlin in the bifurcation structure that makes an unphysical prediction, another comes from a thermal diffusivity term that was regarded as negligible in previous work on this model. A third issue arises from the two strands in literature on the physics of confinement transitions: the model as it stands does not describe confinement transitions due to a nonlinear electric field drive. 

\section{Shear flows also generate turbulence \label{four}}

\begin{figure}[hc]\hspace*{10mm}\begin{minipage}[c]{0.9\textwidth}
    \begin{minipage}[c]{0.58\linewidth}
    \includegraphics[scale=1]{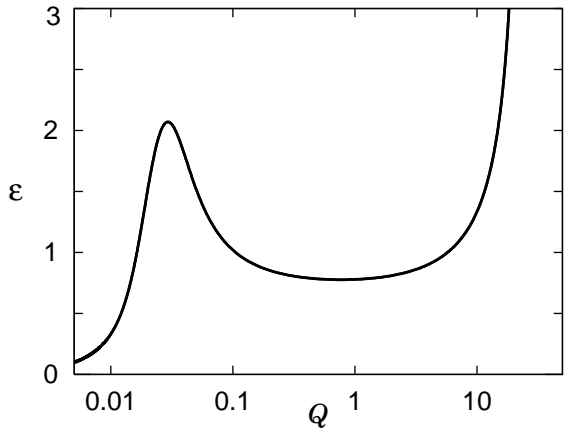}
    \end{minipage} 
    \begin{minipage}[c]{0.38\linewidth}
      \caption{\label{fig5} The curve is the locus of the Hopf bifurcations in Fig. \ref{fig4}(c) over variations in $\varepsilon$.}
    \end{minipage}
 \end{minipage}
\end{figure}

The first issue of incompleteness concerns a pathology in the bifurcation structure of the model, implying infinite growth of shear flow as the power input \textit{falls}. Before we pinpoint the culprit singularity, it is illuminating to evince the physical --- or unphysical --- situation through a study of the role of the thermal capacitance parameter $\varepsilon$, which regulates the contribution of the pressure gradient dynamics, Eq. \ref{e2a}, to the oscillatory dynamics of the system. Conveniently, we can use $\varepsilon$ as a second parameter to examine the stability of steady-state solutions around the Hopf bifurcations in Fig. \ref{fig4} without quantitative change. 
The machinery for this study consists of the real-time equations \ref{e2a}, \ref{e2b}, and \ref{e4} recast in ``stretched time'' $\tau=t/\varepsilon$,
\begin{subequations}\label{e6}
\begin{align}
\frac{dP}{d\tau} & = Q - \gamma NP \label{e6a}\\ 
\frac{dN}{d\tau} & =  \varepsilon\left(\gamma NP - \alpha v^{\prime 2}N - \beta N^2\right) \label{e6b}\\
2\frac{dv^\prime}{d\tau} & =  \varepsilon\left(\alpha v^\prime N - \mu(P,N) v^\prime + \varphi\right),\label{e6c}
\end{align}
\end{subequations}
and the two-parameter locus of Hopf bifurcations in the real-time system shown in Fig. \ref{fig5}. We consider two cases.

\begin{enumerate}
\item The high-capacitance r\'egime: 

The maximum on the curve in Fig. \ref{fig5} marks a degenerate Hopf bifurcation, a DZE point. Here the two Hopf bifurcations at lower $Q$ in Fig. \ref{fig4}(c) merge and are extinguished as $\varepsilon$ increases. (This merger through a DZE has obviously occurred in Fig. \ref{fig4}(d), where $\varphi$ is varied rather than $\varepsilon$.) The surviving upper Hopf bifurcation moves to higher $Q$ as $\varepsilon $ increases further. 

In this high-capacitance r\'egime the dynamics becomes quasi one-dimensional on the stretched timescale. To see this formally, define $\sigma=1/\varepsilon$ and multiply the stretched-time equations \ref{e6b} and \ref{e6c} through by~$\sigma$.  In the limit $\sigma \rightarrow 0$ the kinetic energy variables are slaved to the pressure gradient (or potential energy) dynamics. Switching back to real time and multiplying Eq. \ref{e2a} through by $\sigma$, for $\sigma\rightarrow 0$ $P\approx P_0$. The kinetic energy subsystems see the potential energy as a constant, ``infinite source''. It is conjectured that the surviving upper Hopf bifurcation moves toward $(Q,v^\prime)= (\infty,\infty)$ and for $\varepsilon \gg 1$ the dynamics becomes largely oscillatory in real time, with energy simply sloshing back and forth between the turbulence and the shear flow. 

\item The low-capacitance r\'egime: 

The minimum on the curve in Fig. \ref{fig5} marks another degenerate Hopf bifurcation, also occurring at a DZE point. Here the two Hopf bifurcations at higher $Q$ in Fig. \ref{fig4}(c) merge and are extinguished as $\varepsilon$ decreases. The surviving Hopf bifurcation moves to lower $Q$ --- and higher $v^\prime$ --- as $\varepsilon $  decreases further. This scenario is illustrated in Fig. \ref{trap4a}, where the steady-state curve and limit cycle envelope are roughly sketched in a decreasing~$\varepsilon$ sequence.
\begin{figure}[hc]\hspace*{10mm}
\begin{minipage}[c]{0.9\textwidth}
\includegraphics[scale=0.7]{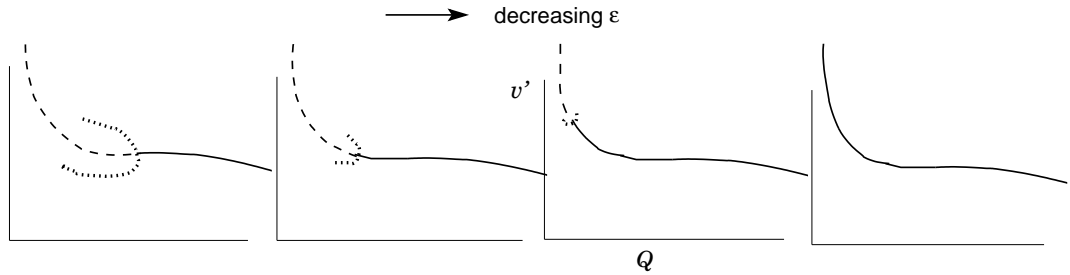}
\caption{\label{trap4a} As $\varepsilon$ is decreased further than the minimum in Fig. \ref{fig5} the remaining Hopf bifurcation slides up the steady-state curve, which becomes stable toward unrealistically high $v^\prime$ and low $Q$.}
\end{minipage}
\end{figure} 

In this low-capacitance r\'egime the dynamics also becomes quasi one-dimensional, and as $\varepsilon \rightarrow 0 $ the conjectured fate of the surviving Hopf bifurcation is a double zero eigenvalue trap at $(Q,v^\prime)= (0,\infty)$. To see why this can be expected, consider again the stretched-time system, Eqs \ref{e6}. For $\varepsilon \ll 1 $ we have $N\approx N_0$ and $v^\prime\approx v_0$. On the stretched timescale the potential energy subsystem sees the kinetic energy subsystems as nearly constant, and $P\approx\left(P_0 - Q/\left(N_0\gamma\right)\right)\exp\left(-N_0\gamma\tau\right) +Q/\left(N_0\gamma\right)$. Reverting to real time, as $\varepsilon dP/dt \rightarrow 0$ we have $P\approx Q/(\gamma N)$; the potential energy is reciprocally slaved to the kinetic energy dynamics. 

The anomaly in this low-capacitance picture is that, as the power input $Q$ ebbs, the shear flow can grow quite unrealistically. 
With diminishing $\varepsilon$ the Hopf bifurcation moves upward along the curve, the branch of limit cycles shrinks, and the conjugate pair of pure imaginary eigenvalues approaches zero. It would seem, therefore, that some important physics is still missing from the model. 

\end{enumerate}

\subsection{A trapped singularity is found and released}

\begin{figure}[ht]
\hspace*{6mm}
\hbox{
\includegraphics[scale=0.9]{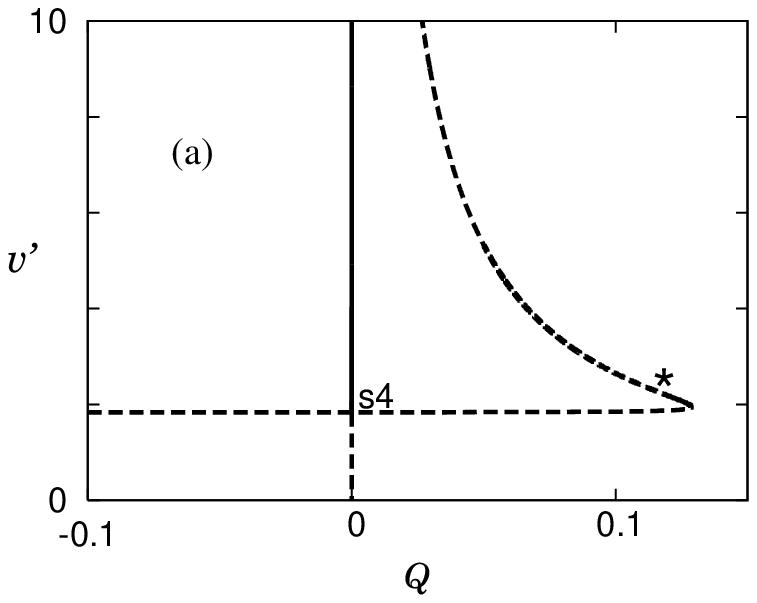}
\includegraphics[scale=0.9]{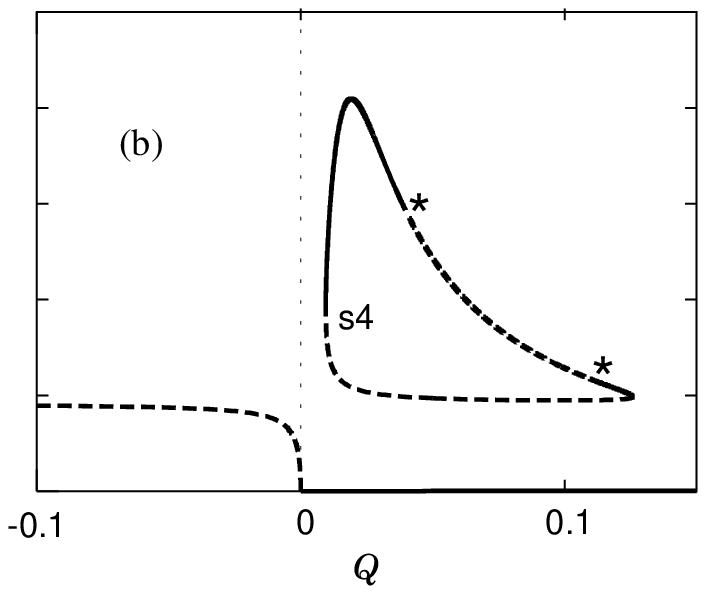}
}
\caption{\label{fig7}(a) $\kappa=0$, (b) $\kappa=0.001$. $\varphi=0.08$, $\beta=0.3$, $b=1$, $a=0.3$, $\alpha=2.4$, $\gamma=1$, $\varepsilon=1$. The starred stability changes are Hopf bifurcations. }
\end{figure}

What is not shown in Figs \ref{fig3} and \ref{fig4}  (because a log scale is used for illustrative purposes) is  a highly degenerate branch of equilibria that exists at $Q=0$ where $N=0$ and $v^\prime=(P^{3/2}\varphi)/b$; it is shown in Fig. \ref{fig7}(a). For $\varphi > 0$ there is a trapped degenerate turning point, annotated as s4, where the ``new'' branch crosses the $Q=0$ branch. 

The key to its release (or unfolding) lies in recognizing that kinetic energy in large-scale structures inevitably feeds the growth of turbulence at smaller scales, as well as vice versa \cite{Terry:2000}. In a flow where Lagrangian fluid elements locally experience a velocity field that is predominantly two-dimensional there will be  a strong tendency to upscale energy transfer (or inverse energy cascade, see Kraichnan and Montgomery \citeyear(1980){Kraichnan:1980}), but the net rate of energy transfer to high wavenumber (or Kolmogorov cascade, see Ball \citeyear(2004){Ball:2004a}) is not negligible. What amounts to an ultraviolet catastrophe in the physics when energy transfer to high wavenumber is neglected maps to a trapped degenerate singularity in the mathematical structure of the model. 
The trapped singularity s4 may be unfolded smoothly by including a  simple, conservative, back-transfer rate between the shear flow and turbulent subsystems:
\begin{subequations}\label{e7}
\begin{align}
\frac{dN}{dt} & =  \gamma NP - \alpha v^{\prime 2}N - \beta N^2+ \kappa v^{\prime 2}\label{e7a}\\
2\frac{dv^\prime}{dt} & =  \alpha v^\prime N - \mu(P,N) v^\prime + \varphi - \kappa v^\prime. \label{e7b}
\end{align}
\end{subequations}
The model now consists of Eqs \ref{e7} and \ref{e2a}, with \ref{e2d}, and the corresponding energy transfer schematic is Fig. \ref{fig1}(c). The back- transfer rate coefficient $\kappa$ need not be identified with any particular animal in the zoo of plasma and fluid instabilities, such as the Kelvin-Helmholtz instability; it is simply a lumped dimensionless parameter that expresses the inevitability of energy transfer to high wavenumber. 

The manner and consequences of release of  the turning point s4 can be appreciated from Fig.~\ref{fig7}(b), from which we learn a salutary lesson: unphysical equilibria and singularities should not be ignored. 
The unfolding of s4 creates a maximum in the shear flow, and
(apparently) a \textit{fourth} Hopf bifurcation is released from a trap at infinity.  At the given values of the other parameters this unfolding of s4 has the effect of forming a finite-area isola of steady-state solutions, but it is important to visualize this (or, indeed, any other) bifurcation diagram as a slice of a three-dimensional surface of steady states, where the third coordinate is another parameter. (Isolas of steady-state solutions were first reported in the chemical engineering literature, where nonlinear dynamical models typically include a thermal or chemical autocatalytic reaction rate \cite{Ball:1999a}.) 

\begin{figure}[hc]
\hbox{
\includegraphics[scale=0.6]{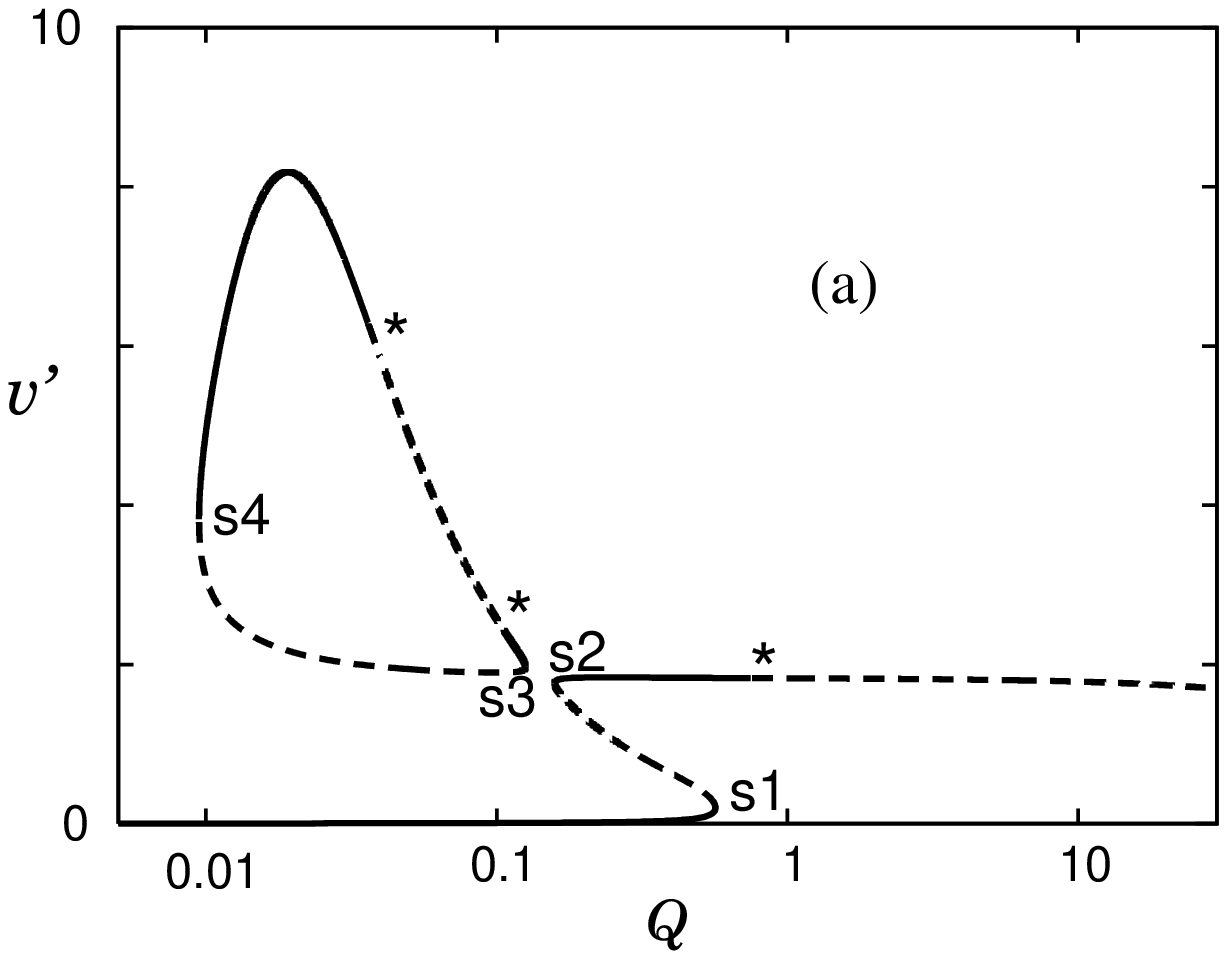}
\includegraphics[scale=0.6]{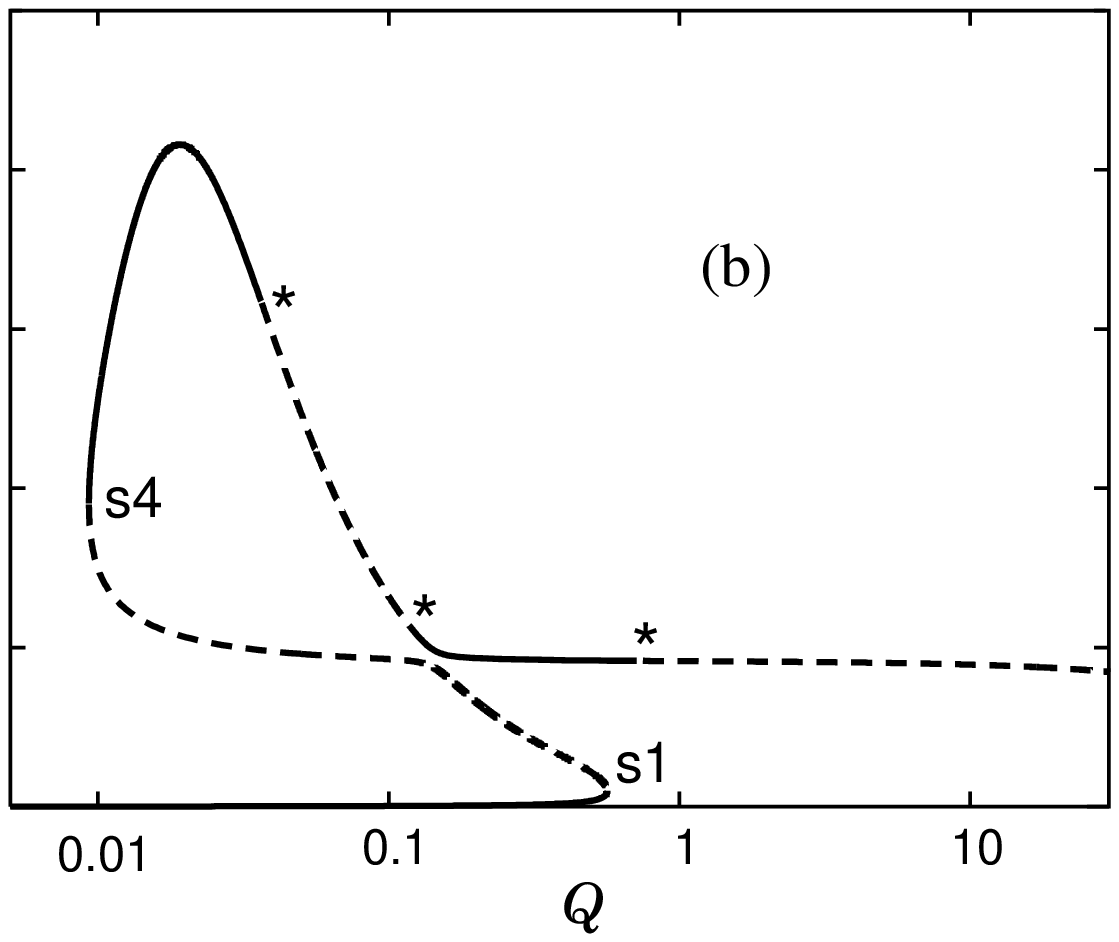}
}
\caption{\label{fig8}  (a)  $\varphi=0.083$, (b) $\varphi=0.084$. $\kappa=0.001$, $\beta=0.3$, $b=1$, $a=0.3$, $\alpha=2.4$, $\gamma=1$, $\varepsilon=1$. Starred points are Hopf bifurcations.}
\end{figure}

In  Fig.~\ref{fig8} we see two slices of this surface, prepared in order to  demonstrate that the metamorphosis identified in section \ref{3.3} is preserved through the unfolding of s4. Here the other turning points are labelled s1, s2, and s3. 
Walking through Fig. \ref{fig8} we make the forward transition at s1 and progress along this branch through the onset of an a limit cycle r\'egime, as in Fig.~\ref{fig3}. For obvious reasons we now designate this segment as the \textit{intermediate} shear flow branch, and the isola or peninsula as the \textit{high} shear flow branch. In (a) a back-transition occurs at s2. The system can only reach a stable attractor on the isola by a transient, either a non-quasistatic jump in a second parameter or an evolution from initial conditions within the appropriate basin of attraction. In (b) as we make our quasistatic way along the intermediate branch with diminishing $Q$  the shear flow begins to grow, then passes through a second oscillatory domain before reaching a maximum and dropping steeply; the back transition in this case occurs at s4.

\section{Thermal dissipation affects the bifurcation structure\label{five}}

In the model so far the only outlet channel for the potential energy is conversion to turbulent kinetic energy, given by the conservative transfer rate $\gamma PN$.
 However, in a driven dissipative system such as a plasma other conduits for gradient potential energy may be significant. 
 
The cross-field thermal diffusivity, a neoclassical transport quantity \cite{Leontovich:1965} is often assumed to be negligible in the strongly-driven turbulent milieu of a tokamak plasma \cite{Sugama:1994a,Sugama:1995,Ball:2002}, but here Eq.~\ref{e2} is modified to include explicitly a linear ``infinite sink'' thermal energy dissipation rate:
\begin{equation}
\varepsilon\frac{dP}{dt}  = Q - \gamma NP - \chi P.\label{e8}
\end{equation} 
Following Thyagaraja et al. \citeyear(1999){Thyagaraja:1999}  $\chi$ is taken as as a lumped dimensionless parameter and the  rate term $\chi P$ as representing all non-turbulent or residual losses such as neoclassical and radiative losses.  The model now consists of Eqs \ref{e7} and \ref{e8}, with \ref{e2d} and the corresponding energy schematic is Fig. \ref{fig1}(d). 

This simple dissipative term has profound effects on the bifurcation structure of the model, and again the best way to appreciate them is through a guided walking tour of the bifurcation diagrams.  
\begin{figure}[Ht]
\vspace*{-4mm}
\hspace*{8mm}\hbox{
\includegraphics[scale=0.84]{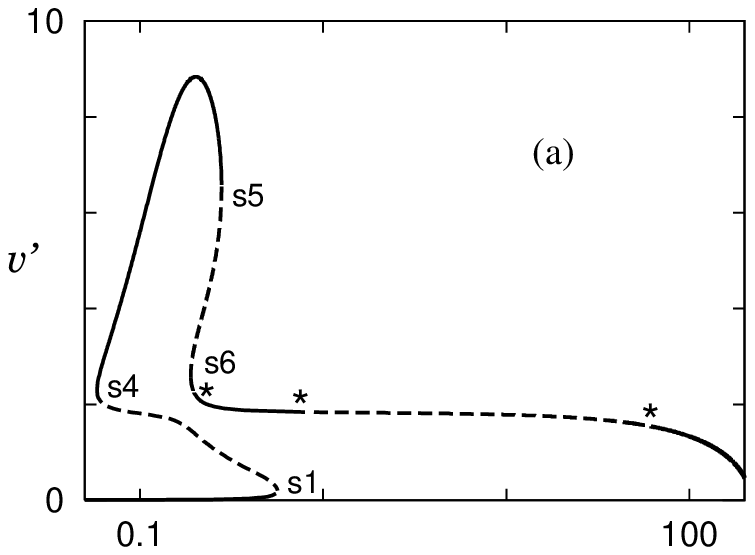}
\includegraphics[scale=0.84]{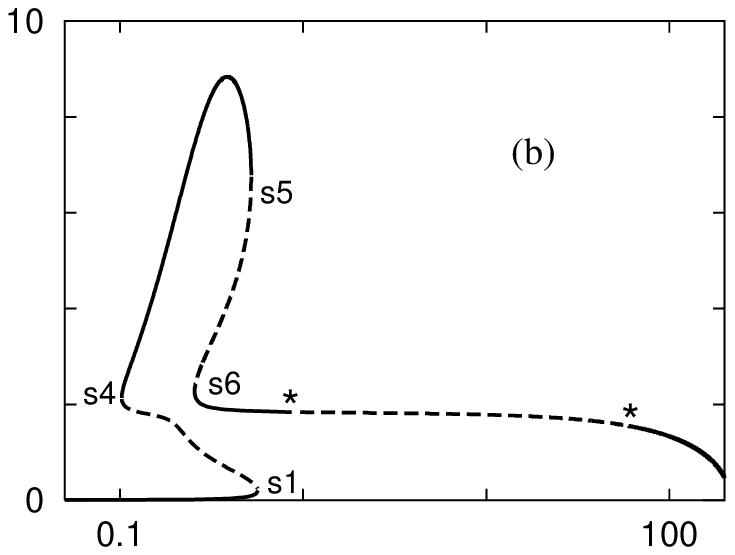}
}
\vspace*{-2mm}
\hspace*{8mm}\hbox{
\includegraphics[scale=0.84]{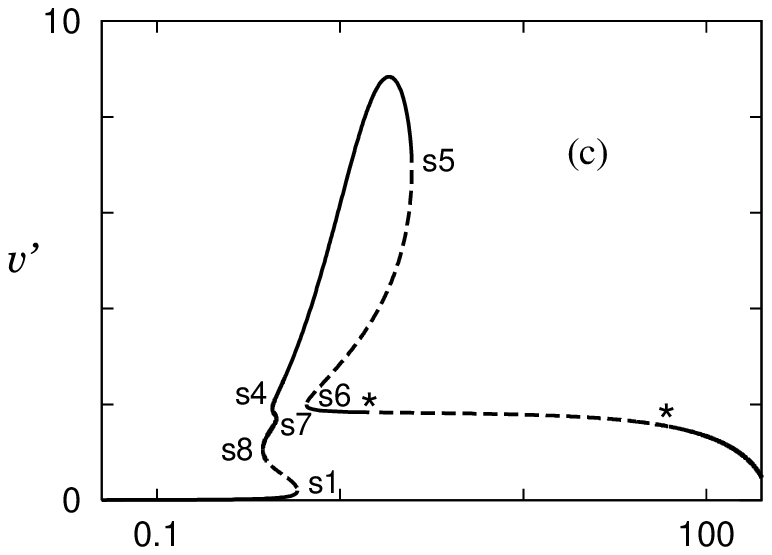}
\includegraphics[scale=0.73]{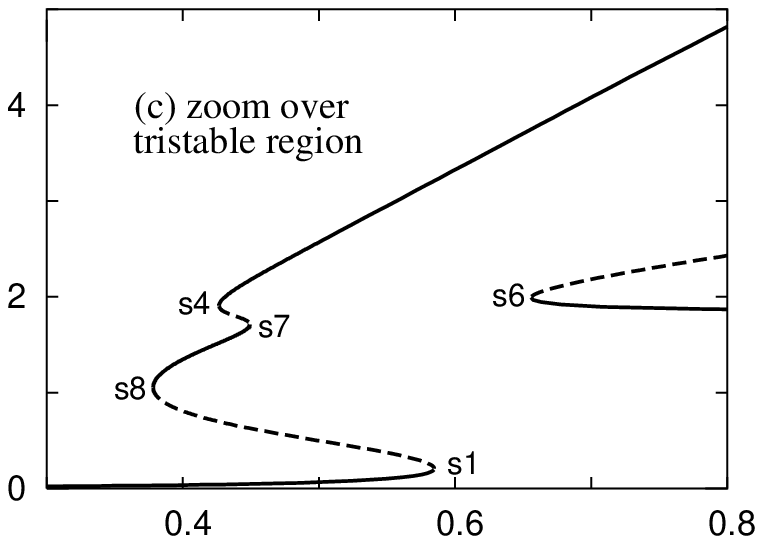}
}
\vspace*{-2mm}
\hspace*{8mm}\hbox{
\includegraphics[scale=0.84]{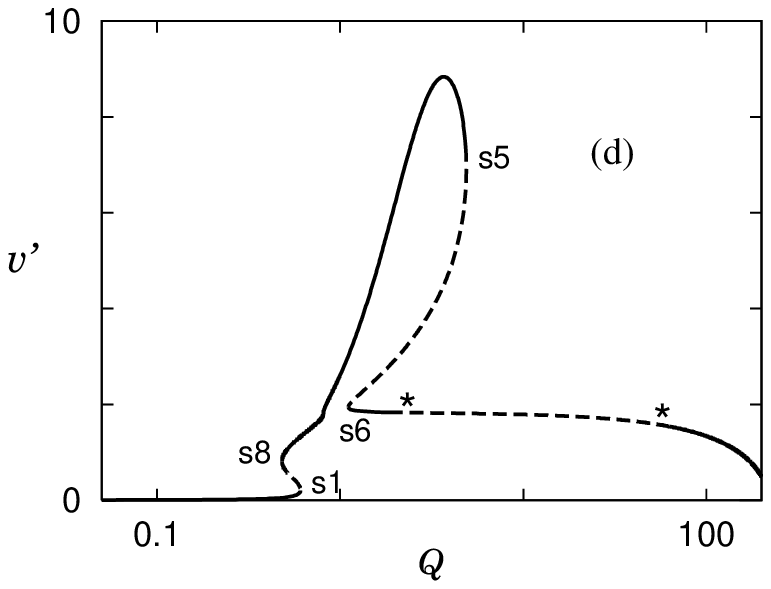}
\includegraphics[scale=0.84]{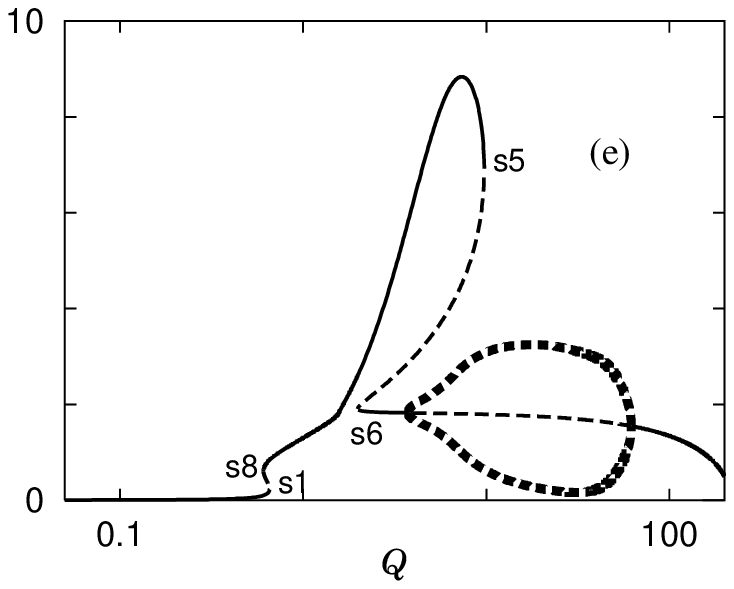}
}
\vspace*{-2mm}
\caption{\label{fig9} (a) $\chi=0.005$, (b) $\chi=0.01$, (c) $\chi=0.05$, (d) $\chi=0.1$, (e) $\chi=0.2$. $\kappa=0.001$, $\varphi=0.088$,  $\beta=0.3$, $b=1$, $a=0.3$, $\alpha=2.4$, $\gamma=1$, $\varepsilon=1$. In (a) to (d) the Hopf bifurcations are starred, and for clarity the limit cycle amplitude envelopes are omitted.}
\end{figure}
In Fig. \ref{fig9} the series of bifurcation diagrams has been computed for increasing values of $\chi$  and a connected slice of the steady state surface (i.e., using a set of values of the other parameters for which the metamorphosis has already occurred). 

A qualitative change is immediately apparent, which has far-reaching consequences: for $\chi>0$ the two new turning points s5 and s6 appear, born from a local cusp singularity that was trapped at $\chi = 0$. Overall, from (a) to (e) we see that s1 does not shift significantly but that the peninsula becomes more tilted and shifts to higher $Q$, but let us begin the walk at s1 in (b).  Here, as in Fig. \ref{fig8},  the transition occurs to an intermediate shear flow state and further increments of $Q$ take the system through an oscillatory r\'egime. But the effect of decreasing $Q$ is radically different: at s6 a discontinuous transition occurs to a  high shear flow state on the stable segment of the peninsula. From this point we may step forward through the shear flow maximum and fall back to the intermediate branch at s5. We see that over the range of $Q$ between s5 and s6 the system has five steady states, comprising three stable interleaved with two unstable steady states. As in Fig. \ref{fig8}(b) a back transition at low $Q$ occurs at s4. 

\begin{figure}[hc]
    \begin{minipage}[c]{0.58\linewidth}
    \includegraphics[scale=1]{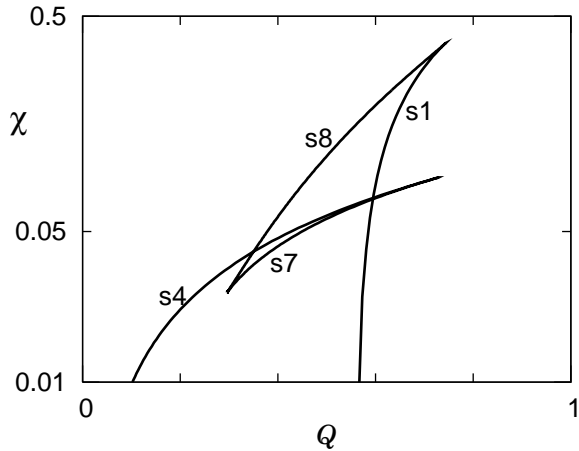}
    \end{minipage} 
    \begin{minipage}[c]{0.38\linewidth}
      \caption{\label{fig9c-2par}The turning points s1, s8, s7, and s4 are followed over  $\chi$. $\kappa=0.001$, $\varphi=0.088$, $\beta=0.3$, $b=1$, $a=0.3$, $\alpha=2.4$, $\gamma=1$, $\varepsilon=1$. }
    \end{minipage}
\end{figure}

The tristable r\'egime in (b) has disappeared in (c) in a surprisingly mundane way: not through a singularity but merely by a shift of  the peninsula toward higher $Q$. But this shift induces a \textit{different} tristable r\'egime through the creation of s7 and s8 at another local cusp singularity. In (d) s4 and s7 have been annihilated at yet another local cusp singularity. 
It is interesting and quite amusing to puzzle over the 2-parameter projection of these turning points s1, s8, s7, and s4 followed over $\chi$, it is given in 
Fig. \ref{fig9c-2par}. The origins of the three local cusps can be read off the diagram, keeping in mind that the crossovers are a \textit{trompe de l'oeil}: they are nonlocal.  

\begin{figure}[hc]
\hspace*{5mm}\hbox{
\includegraphics[scale=0.9]{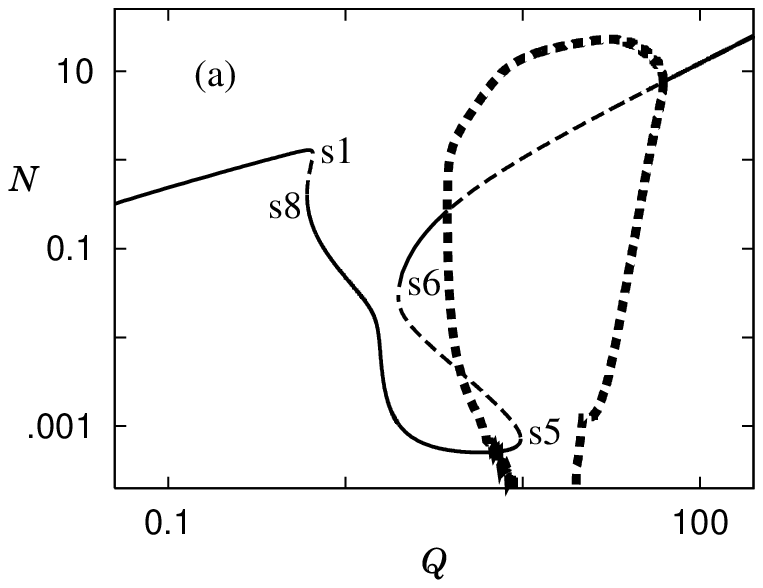}
\includegraphics[scale=0.9]{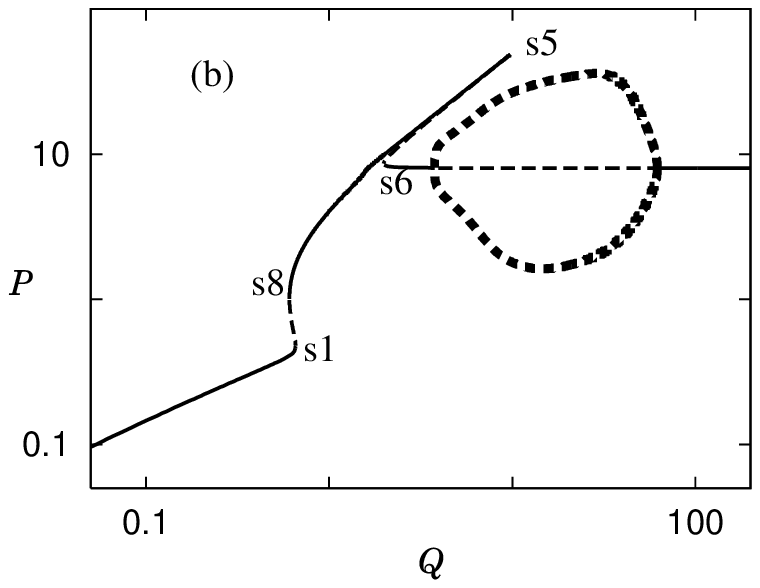}
}
\vspace*{-2mm}
\caption{\label{fig10} The turbulence and pressure gradient subsystems corresponding to Fig. \ref{fig9}(e).}
\end{figure}

At s5 in Fig. \ref{fig9}(c), (d), and (e) the system transits to a limit cycle, rather than to a stable intermediate steady state. Shown in Fig. \ref{fig10} are the bifurcation diagrams in $N$ and $P$ corresponding to Fig. \ref{fig9}(e).  The pressure gradient jumps at s1 because the power input exceeds the distribution rates, and oscillatory dynamics between the energy subsystems sets in abruptly at s5. The turbulence is enormously suppressed due to uptake of energy by the shear flow, but rises again dramatically with this hard onset of oscillations.  

\section{A unified model includes a nonlinear shear flow drive \label{six}}

The early theoretical work on confinement transitions attempted to explain edge L--H transitions  exclusively in terms of the electric field driving torque created by  nonambipolar ion orbit losses  
\cite{Itoh:1988,Shaing:1989,Stringer:1993}, with no coupling to the internal dynamics
of energy transfers from the potential energy reservoir in the pressure gradient. The electric field is bistable, hence the transition to a high shear flow, or high confinement, r\'egime is discontinuous and hysteretic. Although there are many supporting experiments \cite{Fujisawa:2003}, this exposition of the physics behind confinement transitions is incomplete because it cannot explain shear flow suppression of turbulence --- a well-known characteristic of L--H transitions. 

Here this ``electric field bifurcation'' physics is treated as a piece of a more holistic physical picture and  a simple model for the rate of shear flow generation due to this physics is used to create a unified dynamical model for confinement transitions. 
Following the earlier authors this rate is given as
 $r(P) = 
\nu \exp\left[-\left(w^2/P\right)^2\right] $, 
which simply says that the rate at which ions are preferentially lost, and hence flow is generated, is proportional to a collision frequency $\nu$ times the fraction of those collisions that result in ions with sufficient energy to escape. The form of the energy factor assumes an ion distribution that is approximately Maxwellian and $w^2$, analogous to an activation energy, is proportional to the square of the critical escape velocity. In this form of the rate expression I have explicitly included the temperature-dependence of $r$, through $P$, which couples it to the rest of the system.  If $w$ is high the rate is highly temperature (pressure gradient) sensitive. (For heuristic purposes constant density is assumed, constants and numerical factors are normalized to 1, and  the relatively weak temperature dependence of the collision rate $\nu$ is ignored.) 

For convenience the equations for the unified model are gathered together: 
\begin{subequations}\label{e9}
\begin{align}
\varepsilon\frac{dP}{dt} & = Q - \gamma NP  - v^{\prime 2}r(P) - \chi P \label{e9a}\\
\frac{dN}{dt} & =  \gamma NP - \alpha v^{\prime 2}N - \beta N^2+ \kappa v^{\prime 2}\tag{\ref{e7a}}\\
2\frac{dv^\prime}{dt} & =  \alpha v^\prime N - \mu(P,N) v^\prime + v^{\prime }r(P) - \kappa  v^{\prime } \label{e9b}
+ \varphi\\
&r(P)=  \nu \exp\left[-\left(w^2/P\right)^2\right] \label{e8c}\\
&\mu(P,N) =  bP^{-3/2} + a PN. \tag{\ref{e2d}}
\end{align}
\end{subequations}
The corresponding energy schematic is Fig. \ref{fig1}(e) where it is seen that 
$ r(P)$ is a competing potential energy conversion channel, that can dominate the dynamics when the critical escape velocity $w$ is low or the pressure is high. 
\begin{figure}[ht]
\begin{center}
\hbox{
\includegraphics[scale=0.9]{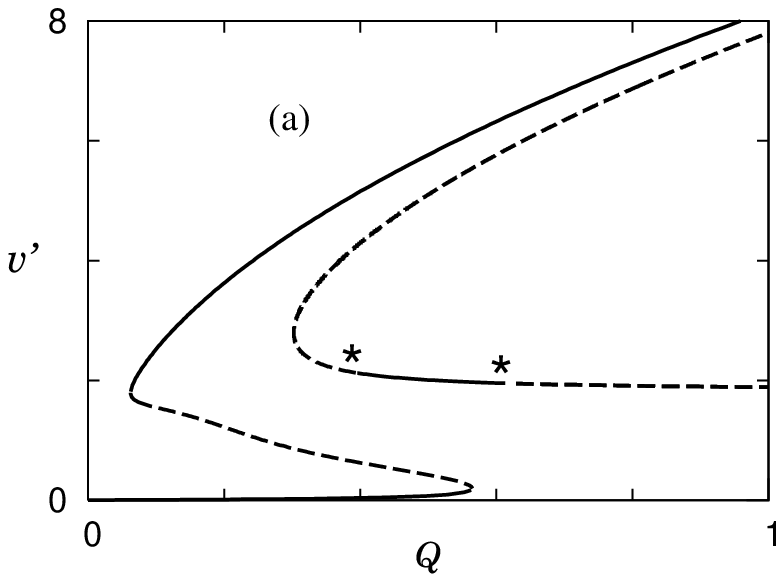}
\includegraphics[scale=0.9]{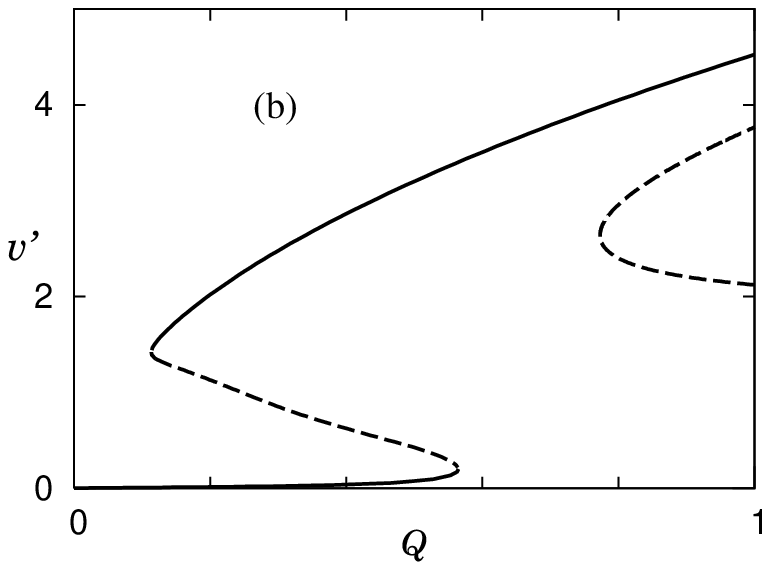}
}
\hbox{
\includegraphics[scale=0.9]{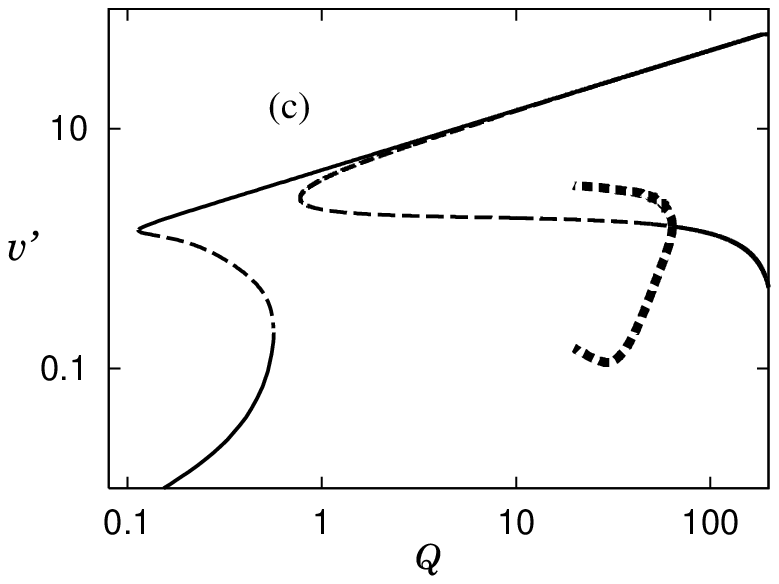}
\includegraphics[scale=0.9]{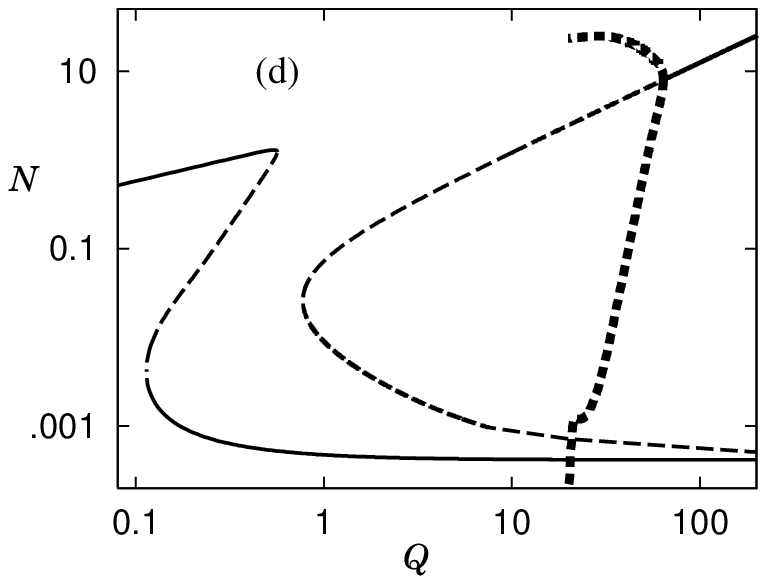}
}
\caption{\label{fig11} (a) $\nu = 0.015$; (b)--(d) $\nu=0.05$. Other parameters: $w=1$, $d=0.01$, $\kappa=0.001$, $\varphi=0.088$, $\beta=0.3$, $b=1$, $a=0.3$, $\alpha=2.4$, $\gamma=1$, $\varepsilon=1$. }
\end{center}
\end{figure}
This is exactly what we see in the bifurcation diagrams, Fig. \ref{fig11}. Overall, the effect of 
this contribution to shear flow generation from the ion orbit loss torque is to elongate and flatten the high shear flow peninsula. The Hopf bifurcations that are starred in (a), where the contribution is relatively small, have disappeared in (b) at a DZE singularity. What this means is that as $r(P)$ begins to take over \textbf{there is no longer a practicably accessible intermediate branch}, as can be seen in (c) where the intermediate branch is unstable until the remaining Hopf bifurcation is encountered at extremely high $Q$. Locally, in the transition region, as  $r(P) $ becomes significant the bifurcation diagram begins to look more like the simple S-shaped, cubic normal form schematics with classical hysteresis presented by earlier authors. However, this \textbf{unified} model accounts for shear flow suppression of the turbulence (d), whereas theirs could not. 

\section{Summary and conclusions \label{seven}}

The generation of stable shear flows in plasmas, and the associated confinement transitions and oscillatory behaviour in tokamaks and stellarators,  is regulated 
by Reynolds stress decorrelation of gradient-driven turbulence
and/or by an induced bistable radial electric field. 
\textbf{These two mechanisms are smoothly
unified  by the first smooth road through the singularity and bifurcation structure of  a reduced dynamical model for this system}. 

 The model is constructed self-consistently, beginning from simple rate-laws derived from the basic pathways for energy transfer from pressure gradient to shear flows. It is iteratively strengthened by finding the singularities and allowing them to ``speak for themselves'', then matching up appropriate physics to the unfoldings of the singularities. 

The smooth road from turbulence driven to electric field driven shear flows crosses interesting territory:
\begin{itemize} 
\item Hysteresis is possible in both r\'egimes and is governed by different physics.
\item A metamorphosis of the dynamics is encountered, near which hysteretic transitions are forbidden. The metamorphosis is a robust organizing centre of codimension 1, even though there are singularities of higher codimension in the system.
\item Oscillatory and tristable domains are encountered. 
\item To travel the smooth road several obstacles are successively negotiated in physically meaningful ways: a pitchfork is dissolved, simultaneously releasing a branch of solutions  from a singular trap at infinity, a singularity is released from a trap at zero power input, and another is released from a trap at zero thermal diffusivity.
\end{itemize}
In particular, these results suggest
strategies for controlling access 
to high confinement states and manipulating oscillatory behaviour in fusion experiments. More generally I have shown that low-dimensional models have a useful role to play in the study of one of the most formidable of complex systems, a strongly driven turbulent plasma. Having survived such a trial-by-ordeal, the methodology is expected to continue to develop as a valuable tool for taming this and other complex systems.

\footnotesize

\begin{table}[ht]\footnotesize\begin{center}
\caption{Glossary of nomenclature}
\label{tab1}
\begin{tabular}{p{2cm}p{8cm}}
\hline
\hline
$P$ & pressure gradient potential energy\\
$N$ & kinetic energy of the turbulence\\
$F$ & shear flow kinetic energy\\
$v^\prime$ & shear flow \\
$\varepsilon$ & dimensionless thermal capacitance\\
$Q$ & power input to the plasma pressure gradient \\
$\gamma$ & turbulence driving rate coefficient\\
$\alpha$ & shear flow driving rate coefficient due to Reynolds stress decorrelation of turbulence\\
$\beta$ & turbulence dissipation rate coefficient\\
$\mu$  &kinematic viscosity \\
$b$ & neoclassical viscosity\\
$a$ & anomalous viscosity \\
$\varphi$ & symmetry-breaking shear flow perturbation\\
$\kappa$ & energy back-transfer (downscale) rate coefficient\\
$\chi$ & thermal diffusivity \\
$\nu$ & ion-ion collision rate\\
$w$ & critical ion escape velocity\\
\hline\hline
\end{tabular}\end{center}
\end{table}

\normalsize

\section*{Acknowledgements}
This work is supported by the Australian Research Council. I thank the referees for helpful comments that have resulted in a better paper, and for their positive endorsements. 

 \newpage
 

\end{document}